\documentclass[10pt,twocolumn,showpacs,longbibliography,amsmath,amssymb,prd,floatfix,superscriptaddress]{revtex4-2}

\usepackage{graphicx}
\usepackage{dcolumn}
\usepackage{bm}
\usepackage{color}
\usepackage{txfonts}
\usepackage{microtype}
\usepackage{epstopdf}
\newcommand\numberthis{\addtocounter{equation}{1}\tag{\theequation}}

\usepackage{epstopdf}

\begin{document}

\title{Fermionic signal of vacuum polarization in strong laser fields}

\author{Ya-Nan Dai}
\affiliation{Department of Physics, Shanghai Normal University, Shanghai 200234, China}
\author{Karen Z. Hatsagortsyan}
\email{k.hatsagortsyan@mpi-hd.mpg.de}
\affiliation{Max-Planck-Institut f\"{u}r Kernphysik, Saupfercheckweg 1,
69117 Heidelberg, Germany}
\author{Christoph H. Keitel}
\affiliation{Max-Planck-Institut f\"{u}r Kernphysik, Saupfercheckweg 1,
69117 Heidelberg, Germany}
\author{Yue-Yue Chen}
\email{yue-yue.chen@shnu.edu.cn}
\affiliation{Department of Physics, Shanghai Normal University, Shanghai 200234, China}
\date {\today}

\begin{abstract}

Vacuum polarization (VP) is investigated for the interaction of a polarized $\gamma$-ray beam of GeV photons with a counterpropagating ultraintense laser pulse. In a conventional setup of a vacuum birefringence measurement, a VP signal is the emerging small circular (linear) polarization of the initially linearly (circularly) polarized probe photons. The pair production via the nonlinear Breit-Wheeler process in such a high-energy environment eliminates part of the $\gamma$-photons in the outgoing $\gamma$-beam, increasing the statistical error and decreasing the accuracy of this VP signal. In contrast, we investigate the conversion of the emerging circular polarization of $\gamma$-photons into longitudinal polarization of the created positrons, considering the latter as the main VP signal. To study the VP effects in the highly nonlinear regime, where the Euler-Heisenberg effective Lagrangian method breaks down, we have developed a Monte-Carlo simulation method, incorporating vacuum birefringence and dichroism via the one-loop QED probabilities in the locally constant field approximation. Our Monte Carlo method will enable the study of VP effects in strong fields of arbitrary configuration. With 10~PW laser systems, we demonstrate the feasibility of detecting the fermionic signal of the VP effect at the 5$\sigma$ confidence level with a few hours of measurement time.

\end{abstract}

\maketitle

\section{Introduction}

Quantum electrodynamics (QED) predicts virtual electron-positron pair creation by a photon in vacuum, resulting in  vacuum polarization (VP) in strong electromagnetic fields and the quantum vacuum behaving as a birefringent medium \cite{Halpern_1933, Weisskopf_1936,berestetskii1982quantum}. This intriguing phenomenon has not been directly proven
in an experiment despite continuous attempts \cite{Ejlli_2020,Cadene_2014,Marklund_2006,di2012extremely}. This is important not only as a proof of nonlinear QED but also 
it may point towards new physics beyond the standard model \cite{Peccei_1977,sikivie1983experimental,schubert2000vacuum,gies2006polarized}.

The vacuum birefringence (VB) signal is enhanced using stronger background fields, longer interaction distances, and a higher probe frequency, and the main hindering factor is the background noise.
The long interaction distance has been implemented in PVLAS \cite{zavattini2012measuring,zavattini2006experimental} and BMV \cite{Cadene_2014} experiments, which aim to measure the ellipticity acquired by a linearly polarized optical light  propagating through a strong static magnetic field ($8.8$ T) of a long extension (1 m), however, without conclusive results so far \cite{Ejlli_2020}.

 The advent of high-intensity optical \cite{Yoon_2021,danson2015high} and x-ray free-electron lasers (XFEL) \cite{geloni2010coherence}, coupled with rapid advancements in x-ray polarimetry (with achievable precision of $8\times 10^{-11}$~\cite{Schulze_2022}), has opened new perspectives for measuring VB with the use of ultrastrong laser fields (with magnetic fields reaching $10^6$ T),  and keV photons of XFELs \cite{DiPiazza_2006,Heinzl_2006,Karbstein_2018,Wistisen_2013,dinu2014vacuum}.
 Using a 10-Petawatt class laser, the induced ellipticity signal can reach up to $ \sim 10^{-4}$ for the XFEL probe \cite{dinu2014vacuum}. The HIBEF consortium is developing the flagship experiment in this regime \cite{HIBEF}.

 Further enhancement of the VB signal is envisaged for a combination of $\gamma$-ray sources \cite{Cantatore_1991} and PW laser facilities \cite{King_2016,Ilderton_2016,nakamiya2017probing,bragin2017high}. The ultrastrong laser fields can also be replaced by the fields of an aligned crystal \cite{apyan2008coherent}. The common VB signal discussed in this setup is the polarization of the $\gamma$-ray beam after the interaction, which relies on the feasibility of sensitive $\gamma$-ray polarimetry, which is a challenging task \cite{nakamiya2017probing}. In the VB setup via laser-$\gamma$-beam collisions copious real pairs are produced due to nonlinear Breit-Wheeler process, which is the source of vacuum dichroism (VD) \cite{bragin2017high}. This effect is especially dramatic when the quantum nonlinearity parameter  is large $\chi_\gamma\gtrsim 1$ \cite{katkov1998electromagnetic}. The pair production decreases the number of $\gamma$-photons in the final state, increasing the statistical error of the VB signal measurement, thus playing the role of undesirable noise.

While in the case of optical and x-ray probes, the treatment of VB is valid within the Euler-Heisenberg effective Lagrangian method, as the probe photon energy is negligible with respect to the electron rest mass,
the QED photon polarization operator in the strong background field should be employed in the case of a $\gamma$-probe. The QED polarization operator within one-loop approximation has been investigated in Refs.~ \cite{Baier_1975, Becker_1975,katkov1998electromagnetic,meuren2013polarization,torgrimsson2021loops}, which has been applied to the VP problem  \cite{bragin2017high,king2023strong}. In particular, in Ref.~\cite{bragin2017high} the feasibility of detecting VB and VD with 10-PW laser systems and GeV $\gamma$-photons on the time scale of a few days was demonstrated. 
{\color{black} For VB in a crystal, circular polarization of $\sim 18\%$ is obtained with incident photons in the energy range of 180~GeV \cite{apyan2008coherent}. Recently, it has been proposed to use helicity flips to detect VB \cite{king2023strong}, however, the obtained signature is of high-order ($\alpha^2$) in the fine structure constant $\alpha$, with a suppressed probability.}

\begin{figure}
    \includegraphics[width=0.48\textwidth]{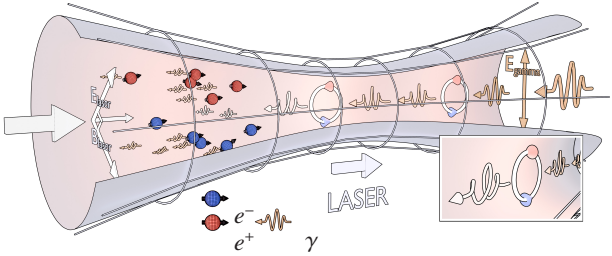}
        \begin{picture}(300,20)
    \put(93,36){$e^-$}
    \put(93,26){$e^+$}
    \put(120,27){$\gamma$}
 \end{picture}
     \caption{Measurement scheme for VP: $\gamma$-photons of linearly polarized penetrate a strong counterpropagating laser pulse, with linearly polarized aligned at 45 degrees with respect to the $\gamma$-polarization. The $\gamma$-photons develop circular polarization due to VB and align along the electric field due to VD. Subsequently, the circular polarization of $\gamma$-photons is transformed into the longitudinal polarization of electrons and positrons as generated in the nonlinear Breit-Wheeler process, yielding a discernible fermionic signal of VP.}
             \label{Fig.scheme}
\end{figure}

In this paper, we put forward a method for observing VB via the created \textit{positron} longitudinal polarization during the interaction of linearly polarized $\gamma$-photons with a linearly polarized ultraintense laser pulse in the highly nonlinear regime with $\chi_\gamma\gtrsim 1$. 
We employ a general scheme of the  pioneering experiment E-144 at SLAC \cite{Bula_1996, Burke_1997,Bamber_1999,Altarelli_2019}, to produce $\gamma$-photons via  Compton scattering and further convert them into electron-positron pairs in an ultrastrong laser field using the nonlinear Breit-Wheeler process. However, we add a polarization perspective to this seminal scheme to exploit it for the application  of a VB measurement. Here the initially linearly polarized $\gamma$-photons propagate in a PW laser pulse, acquiring circular polarization due to VP. The helicity of the photons is subsequently transferred to the produced pairs during the nonlinear Breit-Wheeler process, generating longitudinally polarized positrons with polarization up to $\sim70\%$. Therefore, rather than the conventional \textit{photonic signal} of VP, we find a strong signature of VB in the \textit{positron polarization}, see the scheme of the interaction in Fig. \ref{Fig.scheme}. In contrast to previous schemes where the pair production is undesirable, increasing the statistical error of the VB measurement,  we employ the pairs as a source for a valuable VB signal. To carry out the investigation, we have developed a Monte Carlo method for the simulation of VB and VD of a $\gamma$-ray beam in a highly nonlinear regime, which applies to an arbitrary configuration of a background strong field.
We demonstrate the experimental feasibility of our {\color{black}proposal for} measuring VB with an average statistical significance of 5$\sigma$ on the measurement time scale of a few hours in upcoming 10-PW laser facilities.

\section{vacuum birefringence and dichroism}

Let us first introduce our Monte Carlo method, which allows us to treat the $\gamma$-photon polarization dynamics induced by the VB and VD in strong-field of arbitrary configuration. Until now, the QED Monte Carlo method is known for the simulation of the photon emission and pair production processes \cite{elkina2011qed,ridgers2014modelling,green2015simla,gonoskov2015extended,
chen2019polarized,li2020production,dai2022photon,zhuang2023laser}, which employ the polarization resolved probabilities of the photon emission and pair production in strong fields via the tree diagrams in the locally constant field approximation, see overview in Ref.~\cite{chen2022electron}. The loop diagram contribution of the order of $\alpha$ via the interference  of the one-loop  self-interaction  with the forward scattered one, is also included for the electron, describing the so-called ``no-photon emission" probabilities for the electron polarization change \cite{torgrimsson2021loops,Li_2023}. However, the similar loop diagram contributions for a photon polarization change were missing in the present QED Monte Carlo codes, and have been implemented in this work.

{\color{black} The impact of radiative corrections to photon polarization includes: a polarization generation of $\xi_3$ associated with VD, and a rotation of $\xi_{\perp}=(\xi_1,\xi_2)$ induced by VB, where $\boldsymbol{\xi}_{i}=(\xi_1,\xi_2,\xi_3)$ are the Stokes parameters of the incident photons. 
The former corresponds to the imaginary part of polarization operator, which is related to the pair production probability via the optical theorem, and the latter corresponds to the real part of the polarization operator. The polarization variation of a photon propagating in a background field is described by the Feynman diagrams  shown in Fig. \ref{Fig.diagram}. Panel (a) shows the probability via the tree-level propagation diagram, being zeroth-order in the fine structure constant $\alpha$. Panel (b) presents the probability via the interference diagram of the tree-level propagation diagram and  the one-loop propagation diagram, being first order in $\alpha$. The results of the QED calculations up to the $O\left(\alpha\right)$-order loop contribution \cite{torgrimsson2021loops} are presented in Appendix \ref{A1}. The first term $P^L_{VD}$ of Eq. (\ref{loop}) describes VD, while the second one $P^L_{VB}$ is related to VB. 

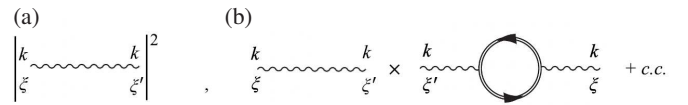
\begin{figure}[b]
    \includegraphics[width=0.5\textwidth]{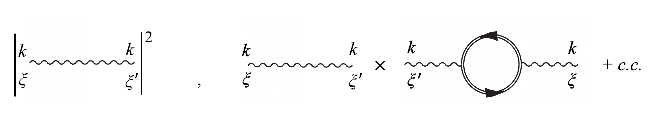}
    \begin{picture}(300,20)
    \put(5,60){(a)}
    \put(85,60){(b)}
 \end{picture}
     \caption{Diagrams contributing to polarization variation of a photon. (a) Zeroth-order in $\alpha$: the tree-level propagation diagram. (b) First order in $\alpha$ : interference diagram of the tree-level propagation diagram and  the one-loop propagation diagram.}    
    \label{Fig.diagram}
\end{figure}

\subsection{Photon polarization due to the  no-pair production probability}

The polarization change due to VD arises because photons with different polarization states are absorbed via pair production differently during propagation. In other words, the dependence of pair production probability by a photon on the photon polarization, will result in the polarization variation of the total photon beam. This selection effect is termed as the change of the photon polarization state during the no-pair production process. We derive below the ``no-pair production" probability, and use it in our modified Monte Carlo code to describe VD.

We begin with the probability for pair production 
\begin{align*}
dW^P & =\frac{\alpha m^{2}d\varepsilon}{\sqrt{3}\pi\omega^{2}}\left\{ \int_{z_{p}}^{\infty}dx\textrm{K}_{\frac{1}{3}}\left(x\right)+\frac{\varepsilon_{+}^{2}+\varepsilon^{2}}{\varepsilon\varepsilon_{+}}\textrm{K}_{\frac{2}{3}}\left(z_{p}\right)-\xi_{i3}\textrm{K}_{\frac{2}{3}}\left(z_{p}\right)\right\},\numberthis
\end{align*}
where $\xi_{i3}$ is the Stokes parameter for linear polarization along polarization basis $\mathbf{\hat{e}}_1=(1,0,0)$ and  $\mathbf{\hat{e}}_2=(0,1,0)$.  The no-pair production probability obtained from the probability conservation is
\begin{align*}
w^{NP}\left(\boldsymbol{\xi}_{i}\right)&=
1-\left\{ \underline{w}+\boldsymbol{\underline{f}}\cdot\boldsymbol{\xi}_{i}\right\} \Delta t,\\
\underline{w}&=\int\frac{\alpha m^{2}d\varepsilon}{\sqrt{3}\pi\omega^{2}}\left[\int_{z_{p}}^{\infty}dx\textrm{K}_{\frac{1}{3}}\left(x\right)+\frac{\varepsilon_{+}^{2}+\varepsilon^{2}}{\varepsilon\varepsilon_{+}}\textrm{K}_{\frac{2}{3}}\left(z_{p}\right)\right],\\
\boldsymbol{\underline{f}}&=-\int\frac{\alpha m^{2}d\varepsilon}{\sqrt{3}\pi\omega^{2}}\hat{e}_{3}\textrm{K}_{\frac{2}{3}}\left(z_{p}\right).\numberthis
\end{align*}
The dependence of pair production probability on photon polarization $\xi_{i3}$ results in a preference of the final polarization state (see also the discussion at Eq. (5.12) in  \cite{yokoya2003user}). Due to this selection effects of initial photon polarization, the final polarization vector after the no-pair production process
becomes
\begin{align*}\label{cain}
\boldsymbol{\xi_{f}^{NP}} & =\frac{\boldsymbol{\xi_{i}}\left(1-\underline{w}\Delta t\right)-\boldsymbol{\underline{f}}\Delta t}{1-\left\{ \underline{w}+\boldsymbol{\underline{f}}\cdot\boldsymbol{\xi_{i}}\right\} \Delta t}=\frac{\boldsymbol{d}^{NP}}{c^{NP}}\numberthis
\end{align*}

We can estimate the polarization variation induced by the no-pair production process as $\Delta\xi_{NP}   =w^{NP}\left(\boldsymbol{\xi}_{i}\right)\left(\boldsymbol{\xi_{f}^{NP}-\xi_{i}}\right)$, and derive the equation for the corresponding  evolution of Stokes parameters:
\begin{align*}
\frac{d\boldsymbol{\xi}_{NP}}{dt} & =\int\frac{\alpha m^{2}d\varepsilon}{\sqrt{3}\pi\omega^{2}}\left(\mathbf{\hat{e}_3}-\left(\boldsymbol{\xi}_{i}\cdot\mathbf{\hat{e}_3}\right)\boldsymbol{\xi}_{i}\right)\textrm{K}_{\frac{2}{3}}\left(z_{p}\right).\numberthis
\end{align*}
Note that, if the photon is in a pure state $\boldsymbol{\xi}_{i}=\pm\mathbf{\hat{e}_3}$,
then there is no polarization variation induced by no-pair production process.
If the photon is in a mixed state along $\mathbf{\hat{e}_3}$ or other directions
other than $\mathbf{\hat{e}_3}$ , then 
\begin{align*}
\frac{d\xi_{1}}{dt} & =-\int\frac{\alpha m^{2}d\varepsilon}{\sqrt{3}\pi\omega^{2}}\xi_{3}\xi_{1}\textrm{K}_{\frac{2}{3}}\left(z_{p}\right),
\end{align*}
\begin{align*}
\frac{d\xi_{2}}{dt} & =-\int\frac{\alpha m^{2}d\varepsilon}{\sqrt{3}\pi\omega^{2}}\xi_{3}\xi_{2}\textrm{K}_{\frac{2}{3}}\left(z_{p}\right),
\end{align*}
\begin{align*}\label{xif}
\frac{d\xi_{3}}{dt} & =\int\frac{\alpha m^{2}d\varepsilon}{\sqrt{3}\pi\omega^{2}}\left(1-\xi_{3}^{2}\right)\textrm{K}_{\frac{2}{3}}\left(z_{p}\right).\numberthis
\end{align*}

\subsection{Vacuum birefringence}

The term $P_{\text{VB}}$ in the loop contribution  is associated with the real part of the polarization operator. It induces a retarded phase between the polarization components along the basis $\mathbf{\hat{e}}_1$ and  $\mathbf{\hat{e}}_2$, resulting in a rotation between $\xi_1$ and $\xi_2$, and in this way contributing to VB. The full VB effect arises due to the net contribution of the $\alpha$-order loop process and the pair-production tree process (with partial cancellation).
In our simulation, the VB is realized by rotation of the  photon polarization vector in $(\xi_1,\xi_2)$ plane at each step \cite{torgrimsson2021loops,bragin2017high,dinu2014vacuum}, see Eq.~(\ref{average}):
\begin{align*}\label{phi}
\left(\begin{array}{c}
\xi_{1}^{f}\\
\xi_{2}^{f}
\end{array}\right) & =\left(\begin{array}{cc}
\cos\varphi & \text{\ensuremath{\sin}}\varphi\\
-\text{\ensuremath{\sin}}\varphi & \cos\varphi
\end{array}\right)\left(\begin{array}{c}
\xi_{1}\\
\xi_{2}
\end{array}\right),\numberthis
\end{align*}
where $\varphi =\frac{\alpha m^{2}}{\omega^{2}}\Delta t\int d\varepsilon\frac{\textrm{Gi}'\left(\xi\right)}{\xi}$, with $\xi=1/ [\delta (1-\delta )\chi_{\gamma} ]^{2/3}$, $\delta=\varepsilon/\omega$, and Gi$'(x)$  the Scorer prime function.

\subsection{Employed Monte-Carlo simulation method for vacuum birefringence and dichroism}

Our modified QED Monte Carlo code is augmented  to include VB and VD via Eqs.~(\ref{cain}) and (\ref{phi}) as described above. Thus, our Monte Carlo method provides the full account for the spin- and polarization-resolved tree-process (nonlinear Breit-Wheeler) and the loop-process (vacuum polarization).
In our Monte Carlo code, at each simulation step $\Delta t$,  the pair production is determined by the total pair production probability, and the positron energy and polarization by the spin-resolved spectral probability  \cite{chen2022electron}, using the common algorithms \cite{zhuang2023laser,dai2022photon,chen2019polarized,li2020production,gonoskov2015extended,
ridgers2014modelling,elkina2011qed,green2015simla}. If the pair production event is rejected, the photon polarization state is determined by the photon-polarization dependent loop probability $w^{NP}$. The full description of the Monte Carlo method is given in Appendix \ref{A2}.

Note that, we are working in the regime of $\chi_\gamma\gtrsim1, \alpha\chi^{2/3}\ll1$, where recoil and pair production are important, but the radiation field is a perturbation. In our simulation, we take into account the $\alpha$-order contributions, i.e. the tree-level first-order processes of photon emission (nonlinear Compton) and pair photoproduction (nonlinear Breit-Wheeler), as well as the one-loop radiative corrections to the electron self-energy (electron mass operator) and photon self-energy (photon polarisation tensor).  The tree-level first-order processes are related to the one-loop self energies by virtue of the optical theorem. In the considered regime, high-order radiative corrections are negligibly small. They become significant only when $\alpha\chi^{2/3}\gtrsim1$ and are therefore not included in our code.

}

\section{The setup for the detection of the vacuum polarization effects in strong laser fields}

\subsection{Generating  a linearly polarized $\gamma-$ray beam via linear Compton scattering}

\begin{figure}[htp]
	\includegraphics[width=0.48\textwidth]{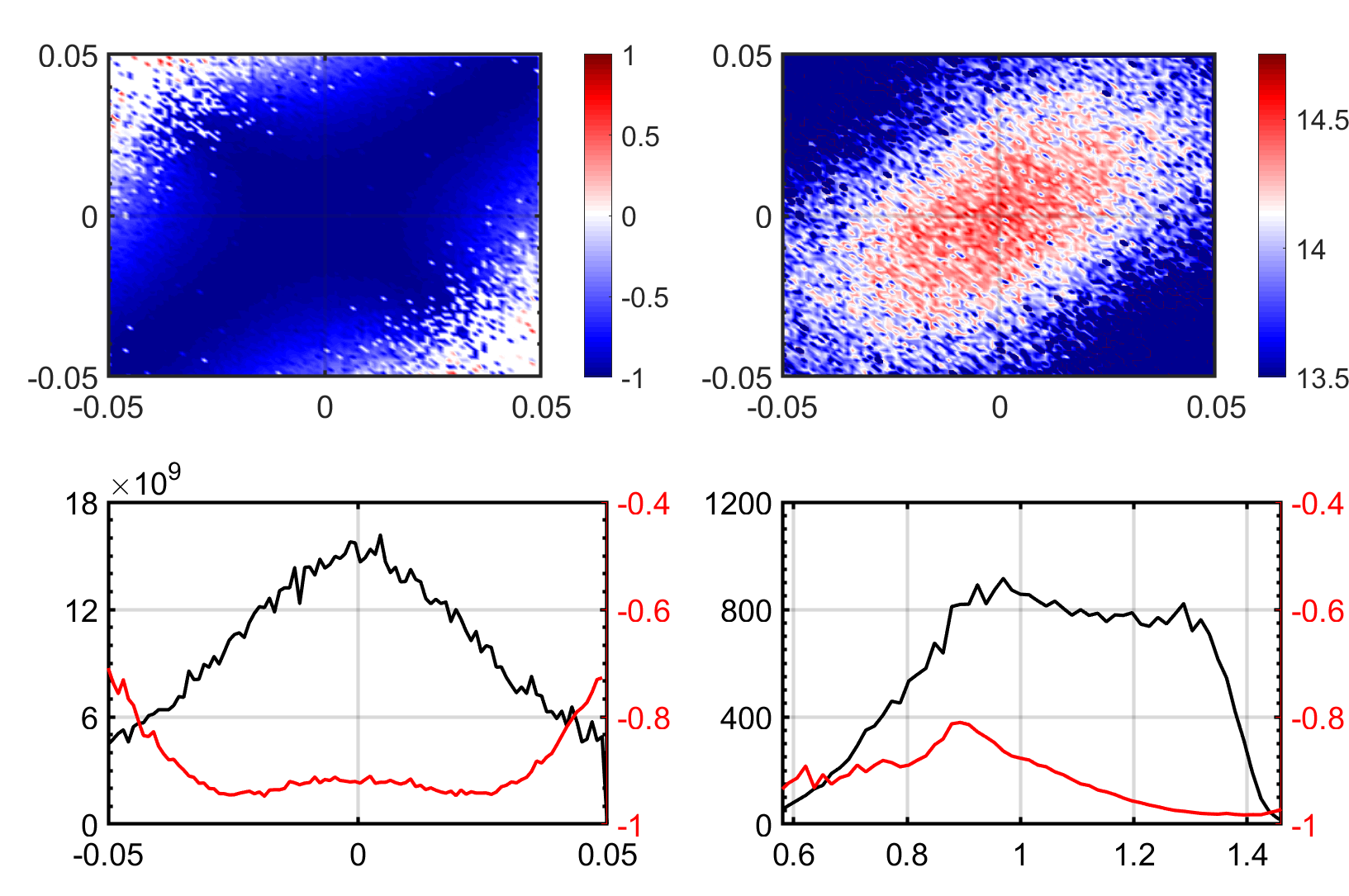}\\
	\begin{picture}(300,20)
    \put(22,164){(a)}
    \put(58,99){\small $\theta_y$}
    \put(2,143) {\rotatebox{90}{\small $ \theta_x$}}
    
	\put(148,164){\color{white}(b)}
	\put(179,99){\small $\theta_y$}
    \put(127,143) {\rotatebox{90}{\small $ \theta_x$}}
    
    \put(23,82){(c)}
    \put(63,15){\small $\theta_x$}
    \put(2,50){\rotatebox{90}{\small $dN_\gamma/d\theta_x$}}
     
	\put(148,82){(d)}
	\put(178,15){\small $\omega$ (GeV)}
	\put(123,47){\rotatebox{90}{\small m$dN_\gamma/d\omega$}}
    \put(242,66){\rotatebox{270}{\small\color{black}$\bar{\xi_1}$}}
	\end{picture}\\
    \caption{(a) Angular distribution of $\gamma$ photon density $\text{log}_{10}d^2N/d\theta_x/d\theta_y$ $ (\text{mrad}^{-2})$ and (b) polarization $\xi_1$ vs $\theta_x$ (mrad) and $\theta_y$ (mrad). (c) The angular distribution of $\gamma$ photon density $dN_\gamma/d\theta_x$ $ (\text{mrad}^{-1})$ (black solid line) and polarization $\xi_1$ (red solid line) vs $\theta_x$.  (d) The energy distribution of $\gamma$ photon density m$dN_\gamma/d\omega$ $ (\text{GeV}^{-1})$ (black solid line) and polarization $\xi_1$ (red solid line) vs $\omega$ $(\text{GeV})$.}
     \label{Fig.gamma}
\end{figure}
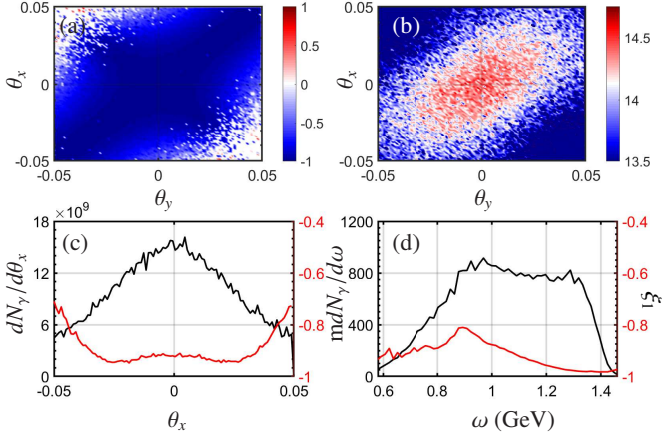
We assume that the probe $\gamma$-photons are  produced by linear Compton scattering of a linearly polarized laser pulse with intensity of $I\sim 10^{16} \text{W/cm}^2$ ($a_0=0.1$) and pulse during of $\tau_p=10$ps. To derive the parameters of the probe $\gamma$-photon beam, we simulate the process with realistic incoming electron beam parameters according to Ref.~\cite{gonsalves2019petawatt,bragin2017high}. The electron beam counterpropagating with the laser pulse consists of $N^0_e=2\times 10^6$ electrons. The electron initial kinetic energy is 8.4GeV, the energy spread {\color{black}$\Delta \varepsilon_0/\varepsilon_0=0.035$}, and the angular divergence $\Delta\theta=0.24\times 10^{-3}$ mrad. The angular distribution and spectrum of emitted photons are obtained using CAIN's code \cite{yokoya2003user}, {\color{black}which takes into account of the electron distribution, angular (energy) divergence of the electron beam, radiation reaction and stochasticity of scattering events.}
The gamma photons within $\theta_{\text{max}}=0.05$mrad are highly polarized with $\overline{\xi}_i=(-0.91,0,0)$,
and have an average energy of $\overline{\omega}_\gamma=1.1$GeV with energy spread {\color{black}$\Delta \omega_{\gamma}/\overline{\omega}_\gamma=0.54$}, see Fig.\ref{Fig.gamma}. The photon yield within $\theta\leq \theta_{\text{max}}$ is $N_\gamma=1\times 10^6\approx 0.5 N^0_{e^-}$. The latter is in accordance with with analytical estimations, see Appendix \ref{A3}. 
{\color{black}The gamma-ray beam can be generated in a beamline similar to LEPS 2 \cite{muramatsu2022spring}, if an upgrade of the laser intensity up to $a_0 = 0.1$, and the electron angular divergence up to $\Delta \theta = 0.24 \times 10^{-3}$ mrad, are implemented.}

\subsection{Fermionic signal of vacuum polarization in strong laser fields}

Afterwards, these photons collide with a 10 PW laser beam for the high-energy VB and VD experiment.
Here we use a 
focused Gaussian linearly polarized laser pulse, with the peak intensity $I\sim 10^{23} \text{W/cm}^2$ ($a_0=150$),  wavelength $\lambda_0=800$~nm, pulse duration $\tau_p=50$~fs,  and the focal radius $w_0=5\lambda_0$  \cite{mckenna2016high,zamfir2014nuclear}.

\begin{figure}[b]
    \includegraphics[width=0.5\textwidth]{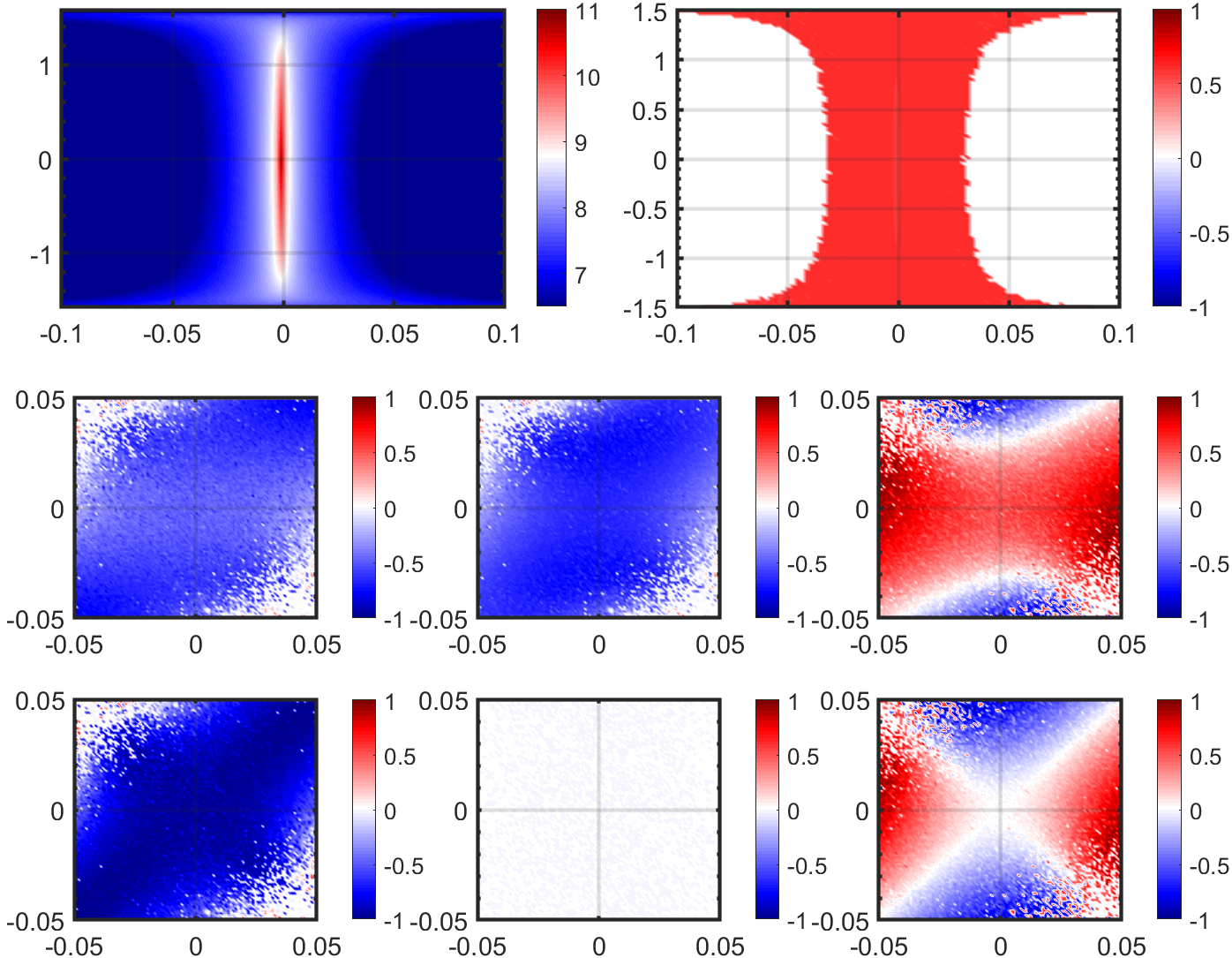}
     \begin{picture}(300,20)
    \put(15,209){ \color{white}(a)}
    \put(50,142){\footnotesize $\theta_y$(rad)}
    \put(-8,173){\rotatebox{90}{\footnotesize$\theta_x$(rad)}}
    \put(145,209){(b)}
    \put(179,142){\footnotesize$\theta_y$(rad)}

    \put(17,130){(c)}
    \put(-8,101){\rotatebox{90}{\footnotesize$\theta_x$(mrad)}}
    \put(101,130){(d)}
    \put(185,130){(e)}

    \put(17,66){(f)}
    \put(28,13){\footnotesize$\theta_y$(mrad)}
    \put(-8,38){\rotatebox{90}{\footnotesize$\theta_x$(mrad)}}
    \put(101,66){(g)}
    \put(112,13){\footnotesize$\theta_y$(mrad)}
    \put(185,66){(h)}
    \put(195,13){\footnotesize$\theta_y$(mrad)}
 \end{picture}

     \caption{ (Top row) The  photon angular distribution after the interaction: (a) for the density $d^2N_\gamma/d\theta_xd\theta_y$, (b) for the photon polarization $\xi_3$,  with $\theta_{x,y}$ in mrad. (Middle row) The angular distribution of photon polarization within $|\theta_{x,y}|\in[0,\theta_{\text{max}}]$ for: (c) degree of linear polarization at $\pm45^\circ$ with respect to polarization basis $P_1^\text{LP}=\xi_1$, (d) degree of circular polarization with $P^\text{CP}=\xi_2$, (e) degree of linear polarization along polarization basis with $P_3^\text{LP}=\xi_3$. (Bottom row) Same as  the middle row, but without VP effects.}
    \label{Fig.pho}
\end{figure}
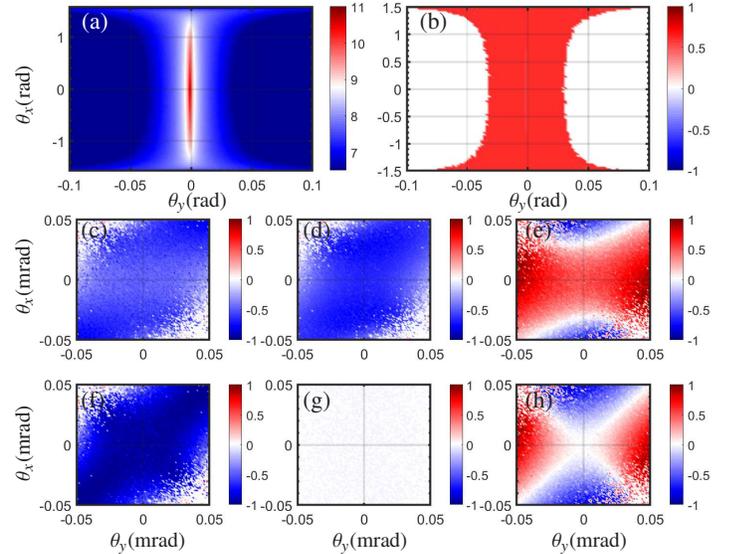

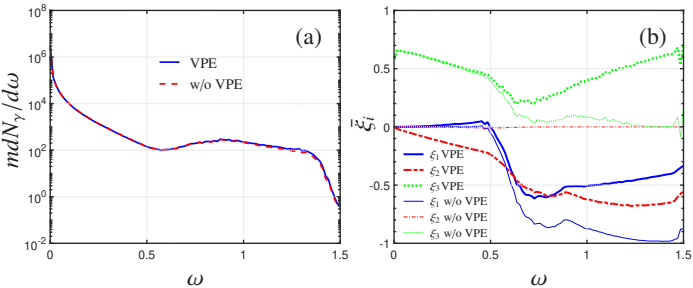
\begin{figure}[b]
    \includegraphics[width=0.5\textwidth]{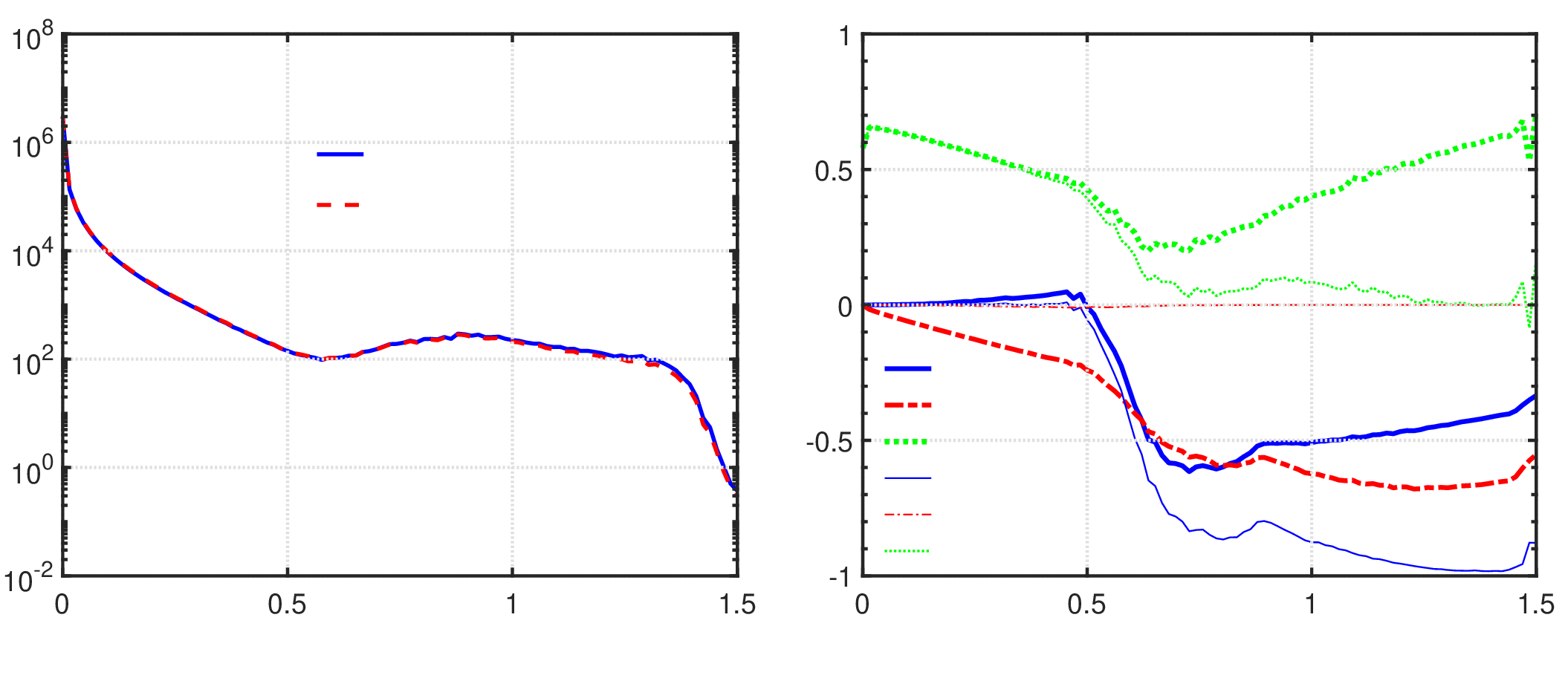}
\begin{picture}(300,20)
    \put(103,113){(a)}
    \put(62,22){$\omega$}
    \put(-8,66){\rotatebox{90}{$mdN_\gamma/d\omega$}}
    \put(63,104){\fontsize{5.5pt}{\baselineskip}\selectfont  VPE}
    \put(63,96){\fontsize{5.5pt}{\baselineskip}\selectfont w/o VPE}

    \put(233,113){(b)}
    \put(192,22){$\omega$}
    \put(124,80){\rotatebox{90}{$\bar{\xi_i}$}}
    \put(154,70){\fontsize{4.5pt}{\baselineskip}\selectfont $\xi_1$VPE}
    \put(154,64){\fontsize{4.5pt}{\baselineskip}\selectfont $\xi_2$VPE}
    \put(154,58){\fontsize{4.5pt}{\baselineskip}\selectfont $\xi_3$VPE}
    \put(154,52){\fontsize{4.5pt}{\baselineskip}\selectfont $\xi_1$ w/o VPE}
    \put(154,46){\fontsize{4.5pt}{\baselineskip}\selectfont $\xi_2$ w/o VPE}
    \put(154,40){\fontsize{4.5pt}{\baselineskip}\selectfont $\xi_3$ w/o VPE}
 \end{picture}
    \caption{(a) Photon spectra with (solid line) and without (red dashed line) VP effect (VPE). (b) The average photon polarization vs photon energy $\omega$ (GeV):  $\xi_1$ (blue solid line), $\xi_2$ (red dot-dashed line), $\xi_3$ (green dotted line), with VP effect (thick line) and without VP effect (thin line).} 
    \label{Fig.pho_1d}
\end{figure}

The simulation results for the final photons are shown in Figs.~\ref{Fig.pho} and \ref{Fig.pho_1d}. The outgoing photon beam consists of the probe photons, survived after pair production ($\sim  10^5$), and a substantial amount of new born photons from radiation of produced pairs ($\sim 10^8$). The remaining probe photons are still confined within $\theta\leq\theta_{\text{max}}$ as the off-forward scattering ($\sim\alpha^2$) is negligible. After propagating through the laser field, the average polarization of probe photons changes to $\overline{\xi}=(-0.53,-0.60,0.37)$  [Fig. \ref{Fig.pho} (c)-(e)], while the larger-angle photons exhibit a distinct linearly polarized: $\overline{\xi}'=(0, 0, 0.59)$ [Fig. \ref{Fig.pho} (b)].

To analyze the simulation results, we use simplified estimations. The VD is described by the following Eq. (\ref{xif}). 
In the case of the photon initial polarization  $\xi_1\approx 1$ and $\xi_3\approx0$, the VD acts as a polarization damper to reduce $\xi_1$ but to increase $\xi_3$. Meanwhile, the VB induces a polarization rotation from $\xi_1$ to $\xi_2$ according to Eq. (\ref{VB_rotate}), resulting in a decrease of $\xi_1$ and an increase of $\xi_2$. With these equations, we estimate the average polarization for a 1~GeV photon after the interaction $\overline{\xi}=(0.53,0.65,0.39)$, which is in a qualitative accordance with  Fig.~\ref{Fig.pho}~(c)-(e).

In the highly nonlinear regime $\chi_\gamma\gtrsim 1$, considerable amounts of 
pairs are produced. The photons emitted by the generated electrons and positrons are mixed with probe photons that carry photonic signals of VP. To clarify the impact of secondary photons and reveal the pure VP effects, we artificially turn off the polarization variation during the no-pair production process. The average polarization of photons at small angle becomes $\overline{\xi}'=(-0.87,0.0,0.06)$, see Fig. \ref{Fig.pho} (f)-(h). The circular polarization $\xi'_2$ disappears without VP regardless of photon emissions.
However, the radiation of pairs affects the linearly polarized of final photons. The average polarization of the emitted photons by unpolarized electrons (positrons), we estimate using the result of Ref.~\cite{dai2022photon}:
\begin{align}\label{xi3}
\xi'_1&=\xi'_2=0,\xi_{3}'=\textrm{K}_{\frac{2}{3}}\left(z_{q}\right)\left[\frac{\varepsilon^{2}+\varepsilon'^{2}}{\varepsilon'\varepsilon}\textrm{K}_{\frac{2}{3}}\left(z_{q}\right)-\int_{z_{q}}^{\infty}dx\textrm{K}_{\frac{1}{3}}\left(x\right)\right]^{-1},
\
\end{align}
where $z_q=\frac{2}{3}\frac{\omega}{\chi_e\varepsilon'}$ with $\varepsilon$ and $\varepsilon'$ being the electron (positron) energy before and after emission, respectively.
Since $\xi'_3$ is inversely proportional to the emitted photon energy $\omega'$, the average polarization at a small angle is reduced by $\sim 1\%$ because of the mixing of the emitted hard photons. For soft photon emissions in the large angle region [Fig.~\ref{Fig.pho} (b)], we have $\xi'_1=\xi'_2=0$ and $\xi'_{3}\approx 0.5$ according to Eq.~(\ref{xi3}), resulting in an average polarization of the entire beam as $\overline{\xi}=(0.0,0.0,0.59)$. In the high-energy regime, photon emission of produced pairs significantly broadens the angular distribution [Fig.~\ref{Fig.pho} (a),(b)], and changes the average polarization of detected photons. Therefore, accounting for the photon emissions is necessary for accurately distinguishing the VP effect.
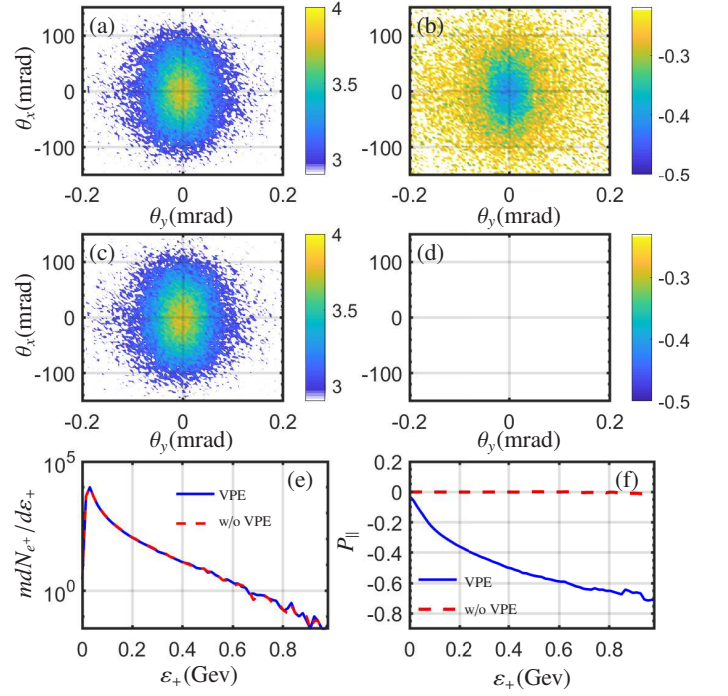
\begin{figure}[t]
    \includegraphics[width=0.5\textwidth]{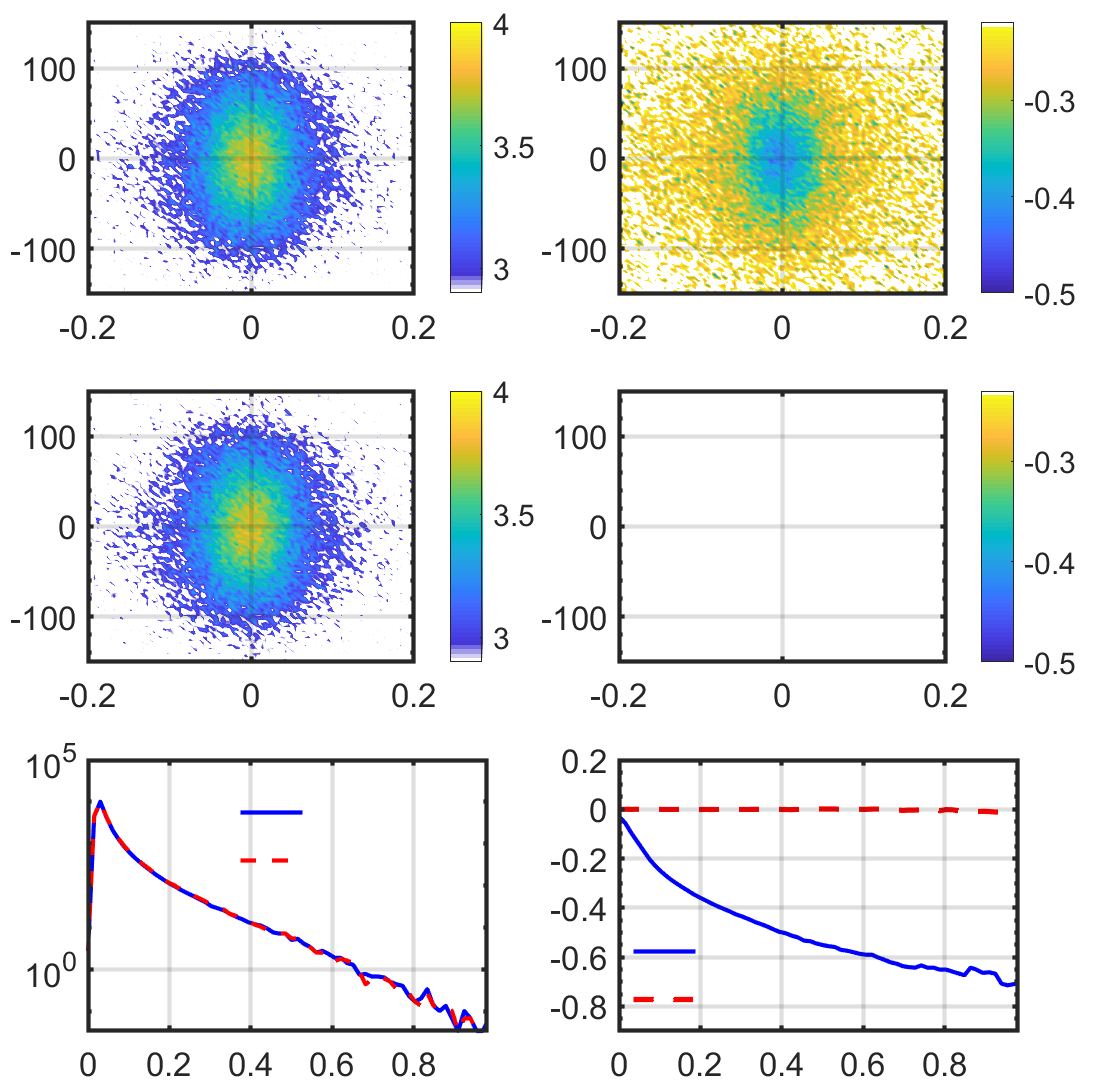}
 \begin{picture}(300,20)
    \put(23,259){(a)}
    \put(46,187){$\theta_y$(mrad)}
    \put(-5,222){\rotatebox{90}{$\theta_x$(mrad)}}
    \put(147,259){(b)}
    \put(170,187){$\theta_y$(mrad)}

    \put(23,173){(c)}
    \put(46,102){$\theta_y$(mrad)}
    \put(-5,136){\rotatebox{90}{$\theta_x$(mrad)}}
    \put(147,173){(d)}
    \put(170,102){$\theta_y$(mrad)}

    \put(98,87){(e)}
    \put(50,13){$\varepsilon_+$(Gev)}
    \put(-5,45){\rotatebox{90}{$mdN_{e^{+}}/d\varepsilon_+$}}
    \put(72,82){\fontsize{5.5pt}{\baselineskip}\selectfont  VPE}
    \put(72,72){\fontsize{5.5pt}{\baselineskip}\selectfont w/o VPE}

    \put(222,87){(f)}
    \put(176,12){$\varepsilon_+$(Gev)}
    \put(117,62){\rotatebox{90}{$P_\parallel$}}
    \put(165,49){\fontsize{5.5pt}{\baselineskip}\selectfont VPE}
    \put(165,39){\fontsize{5.5pt}{\baselineskip}\selectfont  w/o VPE}

 \end{picture}

    \caption{ (Top row) The positron angular distribution: (a) for the number density $d^2N_{e^+}/d\theta_xd\theta_y$ (mrad$^{-2}$), (b) for the longitudinal polarization $P_\parallel$, when $\theta_{x,y}$  are in mrad. (Middle row) Same as top row but without VP effects. (Bottom row) Positron number density $mdN_{e^+}/d\varepsilon_+$  (e), and the longitudinal polarization (f) vs positron energy $\varepsilon_+$ (GeV), with  (blue solid line) VP, and without (red dashed line) VP effect.}
    \label{Fig.pos}
\end{figure}

The full spectrum including all photons is shown in Fig.~\ref{Fig.pho_1d}. The spectrum and polarization exhibit distinct behavior in the two regions divided by $\omega_c=0.6$~GeV. The density distribution in the low-energy region has a feature of synchrotron radiation as it  mostly consists of emitted photons, while the high-energy region exhibits a flat-top structure just as for the probe photons [Fig.~\ref{Fig.pho_1d} (a)]. We find an increase of $\xi_{2,3}$ and decrease of $\xi_1$ in the high-energy region due to VP [Fig.~\ref{Fig.pho_1d}(b)] because the polarization of probe photons is significantly affected by VB and VD. 
Interestingly, the photons emitted in the low-energy region also present a sizeable circular polarization $\xi_2$, indicating that the created $e^+e^-$ pairs obtain longitudinal polarization when taking into account VP.

The polarization features of the created positrons are shown in Fig. \ref{Fig.pos}. The positrons are longitudinally polarized with average polarization of $\sim 13\%$ and highest polarization up to $\sim 70\%$ [Fig.~\ref{Fig.pos}(b) and (f)]. The yield of positrons are $N_{e^+e^-}\approx 7.5\times 10^5\sim 0.75 N_\gamma$ [Fig.~\ref{Fig.pos}(a) and (e)]. In the high-energy region, most of the probe photons are converted to pairs via nonlinear Breit-Wheeler process.
The longitudinal polarization of positrons  stems from the helicity transfer of circular polarization from the probe photons, that is induced by VB at the early stage of interaction.
The emitted photons, detrimental to the high-precision measurement of photonic signals, have a negligible impact on the fermionic signal, as secondary pair production from soft radiation is minimal ($\sim 10^{-2} N_{e^+e^-}$).
Thus, the emergence of longitudinal polarization is essentially a pure signature of VB. As can be seen from Fig. \ref{Fig.pos}(d), the longitudinal polarization vanished without VP.

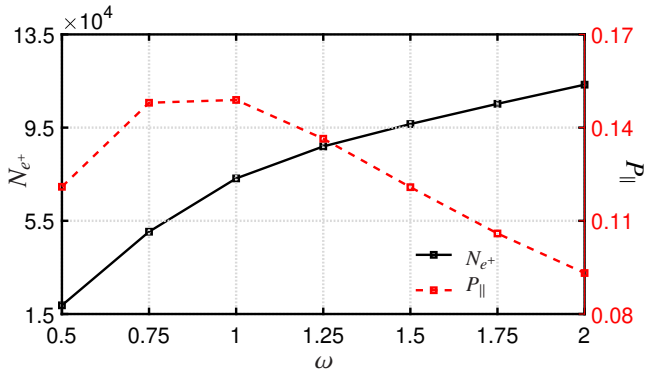
\begin{figure}[b]
    \includegraphics[width=0.5\textwidth]{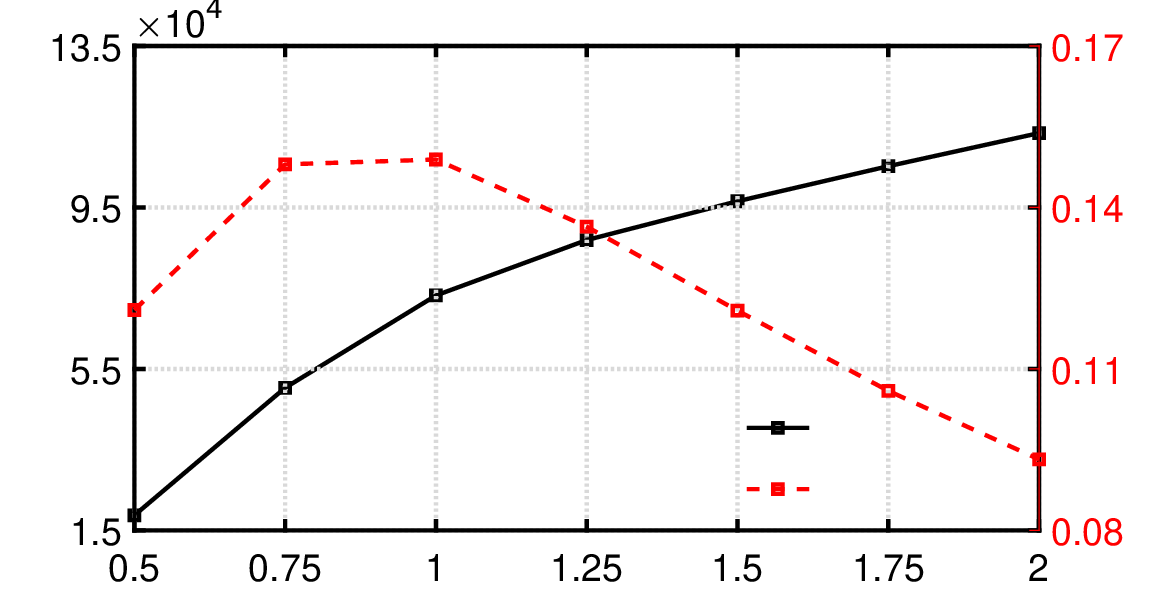}
  \begin{picture}(300,20)
    \put(125,15){\normalsize$\omega$}
    \put(7,83){ \rotatebox{90}{\normalsize  $N_{e^{+}}$}}
    \put(238,94){ \color{black}\rotatebox{270}{\normalsize $P_{\parallel}$}}
    \put(182,54){\fontsize{8pt}{\baselineskip}\selectfont  $N_{e^{+}}$}
    \put(182,44){\fontsize{8pt}{\baselineskip}\selectfont $P_{\parallel}$}
 \end{picture}

    \caption{The scaling law of positron number $N_{e^+}$ (solid black line) and longitudinal polarization $P_\parallel$ (red dashed line) versus energy of probe photon $\omega$ (GeV). The probe photons' number is $N_\gamma=1\times 10^5$ and $\xi=(1,0,0)$. The laser parameters are same as in Figs. \ref{Fig.pho}-\ref{Fig.pos}.}
    \label{Fig.parameter}
\end{figure}

For experimental feasibility, we estimated the impact of probe photons energy on fermionic signals of VB, see Fig.~\ref{Fig.parameter}. As the energy of the probe photon increases, the strength of VB signal also increases due to the larger $\chi_\gamma$, because the high photon energy could accelerate the rotation from $\xi_1$ to $\xi_2$ for a fixed laser duration. Therefore, the longitudinal polarization of positrons increases with photons energy within some limits, before reaching $\sim 15\%$ at $\omega=0.75$~GeV in the case of parameters of Fig.~\ref{Fig.parameter}.
Afterwards, the polarization saturates within some photon energy range, and further decreases with higher $\omega$. This is because with higher $\omega$, and higher $\chi_\gamma$, the probe  photon undergoes pair production before attaining a significant circular polarization due to VB. As a result, unlike the scaling law of positron density that monotonously increases with photon energy, the polarization purity has an optimal energy range within the interval of $\omega/\text{GeV}\in[0.75,1]$.

\section{Experimental feasibility of a vacuum polarization measurement}

\subsection{M{\o}ller polarimetry for detecting positron polarization}

Let us discuss the feasibility of VB detection taking advantage of the positron polarization. There are  conventional techniques for measuring longitudinal polarization of positrons (electrons), such as Compton 
\cite{beckmann2002longitudinal,passchier1998compton,escoffier2005accurate} and M{\o}ller polarimetries \cite{jones2022accurate,arrington1992variable,moller1932theorie}.  For the discussed parameter regime the M{\o}ller polarimetry is more advantageous, which employs the scattering of polarized solid targets off the positrons (electrons)  off a solid targets. Here the longitudinal  polarization is deduced via the measured asymmetry 
$\left\langle R\right\rangle=\frac{N_{+}-N_{-}}{N_{+}+N_{-}}$, where $N_\pm$ are the number of scattered positrons when the positron helicity is parallel or anti-parallel to the target polarization \cite{arrington1992variable}. {\color{black} The cross-section in the center of the momentum frame of the electron reads:
\begin{eqnarray}
\frac{d\sigma}{d\Omega'}= \frac{d\sigma_0}{d\Omega'}\left( 1+\sum_{i,j} P_B^i A_{i,j} P_T^j \right),
\label{sigma}
\end{eqnarray}
where $P_B^i (P_T^j)$ are the components of the beam (target) polarization, as measured in the rest frame of the
beam (target) positrons. Here, we set a new coordinate system with $z'$-axis along the momentum of the positron beam,  and the $y'$-axis normal to the M{\o}ller scattering plane. The prime in the positron coordinate definition is for distinguishing it from that used for the laser-electron interaction.

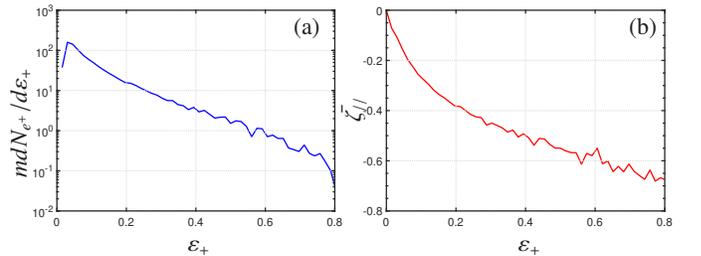
\begin{figure}[b]
	\includegraphics[width=0.48\textwidth]{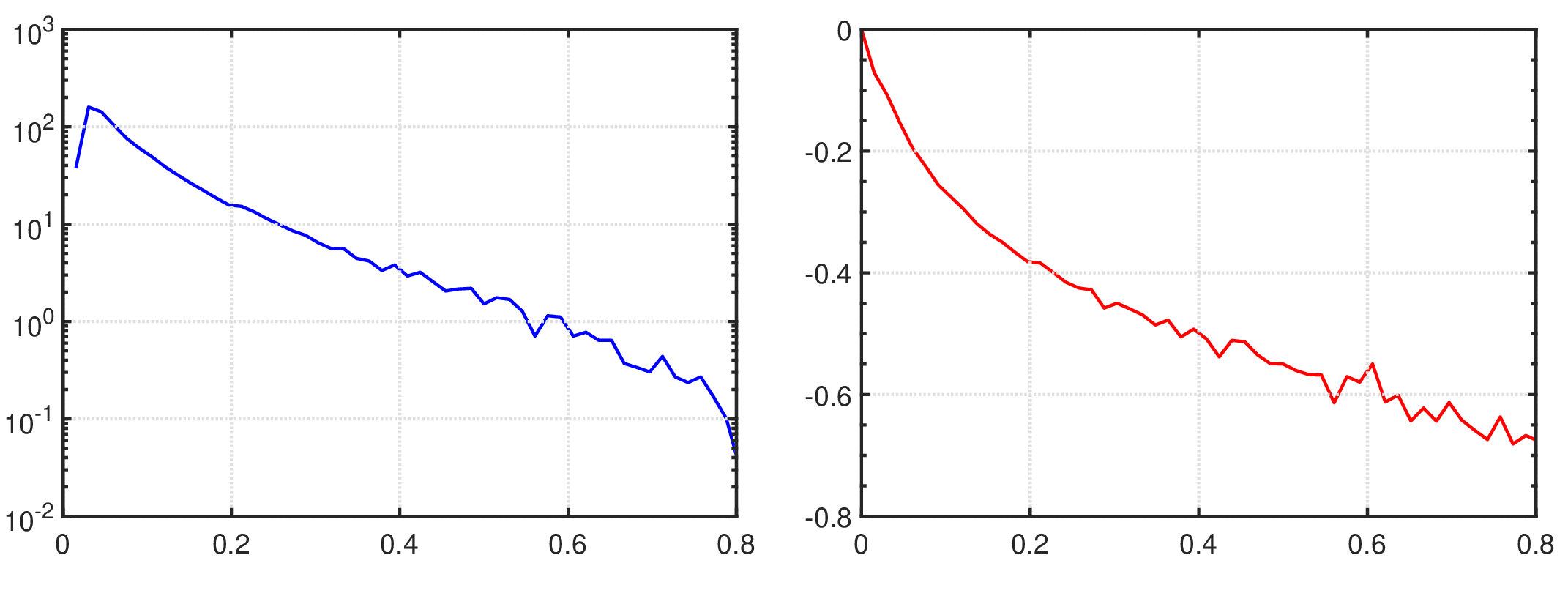}\\
 \begin{picture}(300,20)    
    \put(100,102){(a)}
    \put(60,20){$\varepsilon_+$}
    \put(-7,48){\rotatebox{90}{$mdN_{e^+}/d\varepsilon_+$}}
    \put(227,102){(b)}
    \put(185,20){$\varepsilon_+$}
    \put(118,69){\rotatebox{90}{$\bar{\zeta_{//}}$}} 
 \end{picture}\\
    \caption{ Positron density (a) and longitudinal polarization (b) within 10~mrad.}
       \label{Fig.posndw10mrad}
\end{figure}

The cross section is characterized by the unpolarized cross section $\frac{d\sigma_0}{d\Omega'}$, and nine
asymmetries $A_{i,j}$. The beam polarization components $P_B^i$ are extracted from the measurement of the spin-dependent cross-section on a target of known polarization $P_T$, and using Eq. ~(\ref{sigma}).
To lowest order in QED, the  unpolarized cross-section and nine asymmetries are the following in the ultrarelativistic approximation \cite{arrington1992variable}:
\begin{eqnarray}
\frac{d\sigma_{0}}{d\Omega'}&=&\left[\frac{\alpha\left(1+\cos\theta'_{\text{CM}}\right)\left(3+\cos^{2}\theta'_{\text{CM}}\right)}{2m\sin^{2}\theta'_{\text{CM}}}\right]^{2},\\\nonumber
A_{z'z'}&=-&\frac{(7+{\rm cos}^2 \theta'_{\rm CM}) {\rm sin}^2 \theta'_{CM}}{(3+{\rm cos}^2 \theta'_{\rm CM})^2},\\\nonumber
-A_{x'x'}&=&A_{y'y'}=\frac{{\rm sin}^4 \theta'_{\rm CM} }{(3+{\rm cos}^2 \theta'_{\rm CM})^2},\\\nonumber
A_{x'z'}&=&A_{z'x'}=-\frac{2{\rm sin}^3 \theta'_{\rm CM}{\rm cos} \theta'_{\rm CM} }{\gamma (3+{\rm cos}^2 \theta'_{\rm CM})^2},\\\nonumber
A_{x'y'}&=&A_{y'x'}=A_{y'z'}=A_{z'y'}=0.
\end{eqnarray}
Note that $\theta'_{\rm CM}$ is the center of mass (CM) scattering angle. To measure the longitudinal polarization,  the experimentally determined quantity is the asymmetry parameter 
\begin{eqnarray}
{R}&=&\frac{N_{+}-N_{-}}{N_{+}+N_{-}}.
\label{asy}
\end{eqnarray}
Considering the connection between the Lab scattering angle and the center of mass scattering angle,
 \begin{eqnarray}\nonumber
 {\theta_{\rm L}^{'2} } = 2m_e \left(\frac{1}{p_s}- \frac{1}{p_i}\right),\\
 p_s=\frac{p_i}{2}(1+{\rm cos} \theta'_{\rm CM}),
\end{eqnarray}
the $A_{z'z'}$ is a function of the incident electron energy $\gamma$ and  the detection angle $\theta'_d$ in the Lab frame. Here, $p_s$ ($p_i$) is the momentum of the scattered (incident) positrons for M{\o}ller scattering.

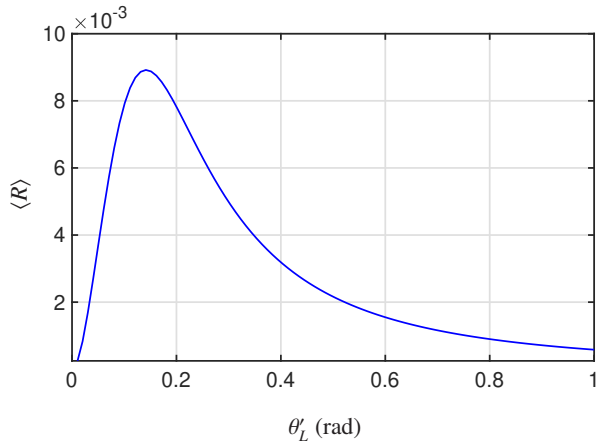
\begin{figure}[b]
	\includegraphics[width=0.5\textwidth]{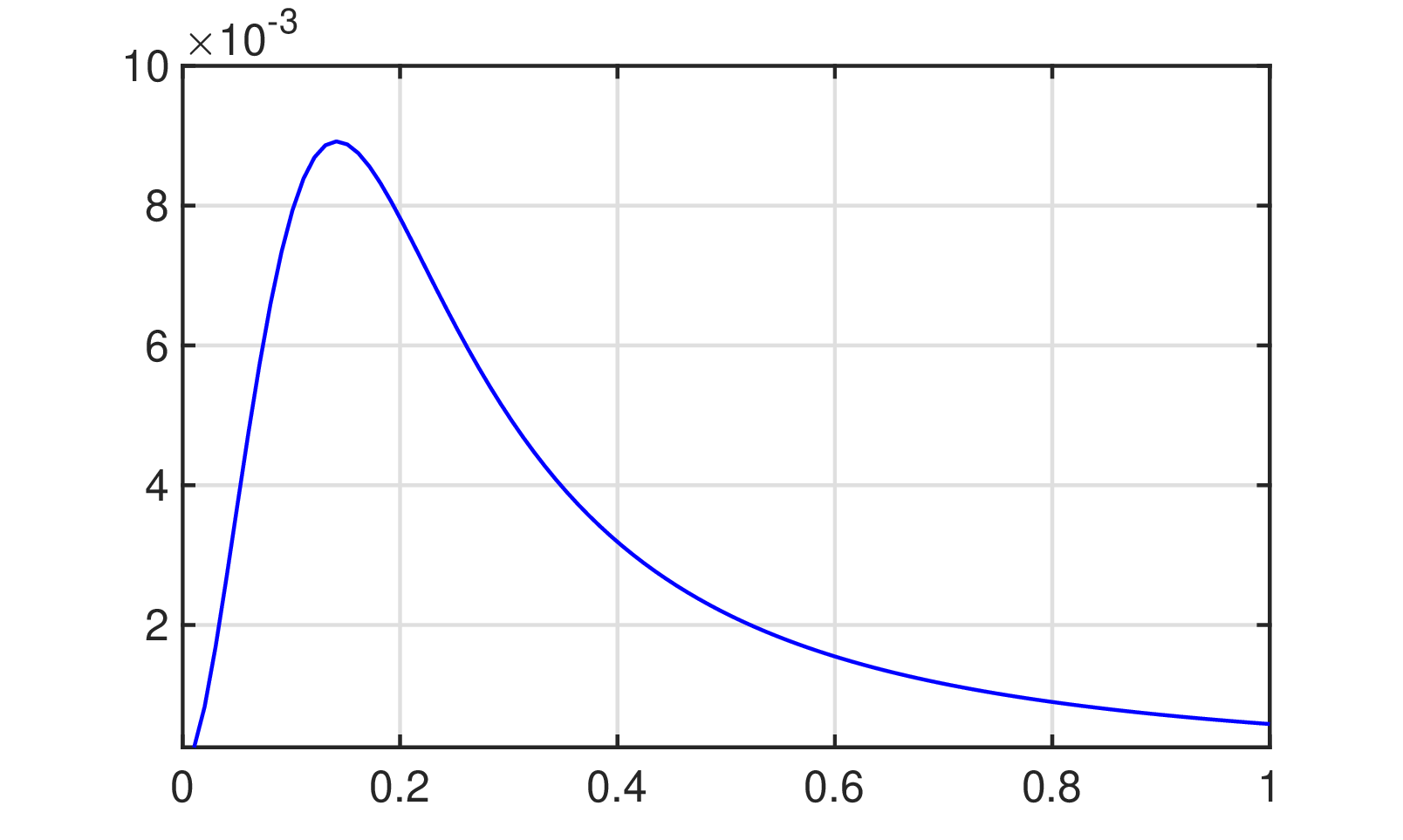}\\	
	\begin{picture}(300,20) 
	 \put(10,95){\rotatebox{90}{$\left\langle R\right\rangle $}}
	 \put(115,10){$\theta_L'$ (rad)}
	\end{picture}\\
	 \caption{The scaling law of asymmetry $\left\langle R\right\rangle$ versus detection angle $\theta_L'$ in the Lab frame.}
\end{figure}

In our setup, the positrons after the interaction are distributed in a wide angle range of $\Delta \theta_x\sim 200$~mrad. We collect the positrons within 10~mrad for the measurement of vacuum birefringence. The spectrum and polarization distribution for positrons within 10~mrad are shown in Fig. \ref{Fig.posndw10mrad}. It can be seen that the positrons have a quite large energy range $\Delta \varepsilon \sim \varepsilon_0$ around the mean energy $\varepsilon_0$. Then, we have to take into account that the rest frames of the particles are different at different energies. In this case, the asymmetry parameter for a certain detection angle $\theta'_d$  is given by:
\begin{align*}
\left\langle R\right\rangle &=\frac{N_{+}-N_{-}}{N_{+}+N_{-}}=\frac{\sum_{i}\sigma_{0i}\zeta_{i}^{z}A'_{zzi}P_{T}^{z}N_{e_{i}^{+}}n_{z}l}{\sum_{i}\sigma_{0i}N_{e_{i}^{+}}n_{z}l}\\
&=\frac{\sum_{i}\sigma_{0i}\zeta_{i}^{z}A'_{zzi}N_{e_{i}^{+}}}{\sum_{i}\sigma_{0i}N_{e_{i}^{+}}}P_{T}^{z},\numberthis
\end{align*}
where $\sigma_{0i}\approx \frac{d\sigma_{0i}}{d\Omega'}\Delta\Omega'$ is the unpolarized cross-section for positrons with energy $\varepsilon^+_i$, and $\Delta\Omega'$ is related to the detection angle in the Lab frame via $\Delta \Omega'=-\frac{8mp_{i}\theta'_{L}}{\left(2m+p_{i}\theta_L^{'2}\right)^{2}}\Delta\theta'_{L}$. $n_{z}$ and $l$ are the density and length of the target, and $N_{e_i^{+}}$ is the number of positrons with energy $\varepsilon^+_{i}$, respectively. The maximum target polarization is $P_{T}^{z}=8.52 \%$. The maximum of asymmetry is $\left\langle R\right\rangle _{max}\approx0.0089$ for $\theta'_{L}=0.1414$~rad. The current experimental capability of measuring the asymmetry parameter is $A_{m}=0.5\%\times P_{T}\times\frac{7}{9}=3.89\times10^{-3}P_{T}\ll \left\langle R\right\rangle _{max}=0.1\times P_{T}$.

Next, we estimate the measurement time for vacuum birefringence with $5\sigma$ confidence level.
The thin foil circular disks used in the M{\o}ller polarimeter are a few microns thick (13~$\mu$m$-$25$~\mu$m), which should be smaller than the milliradiation length (mRL=$10^{-3}$ radiation length) to avoid secondary photon emissions.  Consider a target composed of a Fe-Vo alloy (Supermendur: 49\% Fe, 49\%Co, 2\%Va by mass). A 25~$\mu m$ foil is only 
$ 1.5$~mRL, which can be used in a M{\o}ller polarimeter. The density of the target is 8.12 g/cm$^{3}$ [Density
near r.t. (g/cm$^{3}$): Fe 7.874, Co 8.9, Ni 8.902]. The electron density of the target can be calculated from $n_{z}=\rho N_{A}\left\langle Z\right\rangle /\left\langle A\right\rangle\approx2.26\times10^{24}$
where $\left\langle Z\right\rangle $ and $\left\langle A\right\rangle $
are the average atomic number and mass number of the Vacoflux alloy,
$N_{A}=6.022\times10^{23}$ is the Avogadro's number, $\rho$ is the
density of the foil. Average atomic number $\left\langle Z\right\rangle $ is
26.43 (Fe-26, Co- 27, Va-23), and average mass number 57~a.u. (Fe-55.8~a.u.,
Co- 58.9~a.u., Va-50.94~a.u.). 

The standard deviation then can be estimated with 
\begin{align}
\text{\ensuremath{\Delta R}}&=\frac{1}{\sqrt{N_{+}+N_{-}}}=\frac{1}{\sqrt{\sum_{i}\sigma_{0i}\left(\theta'_{L}=0.14\right)N_{e_{i}^{+}}n_{z}l}}.
\end{align}

For the detecting angle of $\theta_{L}'=0.14$~rad with $\Delta \theta_{L}'=0.03$~rad, the standard deviation is $\Delta R=0.0236$. To achieve a confidence level of $\left\langle R\right\rangle =5\text{\ensuremath{\Delta R}}$, one needs $\tilde{N}_{e^{+}}=2.35\times10^{8}$ positrons. Assuming electron bunches with $N_{e^-}^0=1\times 10^8$ is used for Compton backscattering, our scheme could generate  $N_{e^+}=1.3\times10^6$ positrons within 10~mrad. Using a few PW laser with a repetition rate of 1/60 Hz \cite{mckenna2016high,negoita2022laser}, the measurement of vacuum birefringence with $5\sigma$ confidence level requires a measurement time of $\tilde{N}_{e^{+}}/N_{e}/(1/60)/3600\approx3$ hours.  The measurement time can be further reduced if all  outgoing positrons are focused to a small angle and included in the measurement.

Achieving a $5\sigma$ confidence level for fermionic signals requires 180 shots of a 10-PW laser. For a laser with a repetition rate of 1/60 Hz, it requires 3 hours in a continuous measurement. In real experiments, the  measurement is still feasible but with an extended measurement time to maintain the quality of each laser shot. For instance, in the SULF-10 PW beamline, completing 180 shots of a 10-PW laser usually takes approximately 2 months \cite{li2022acceleration}.
Meanwhile, the measurement time can be reduced at the expense of confidence level. A measurement time of 7 minutes is implied for a measurement with $\sigma$ confidence level.

The estimation of the M{\"u}ller polarimetry signal is given in Appendix \ref{A4}.

\begin{figure}[t]
 \includegraphics[width=0.48\textwidth]{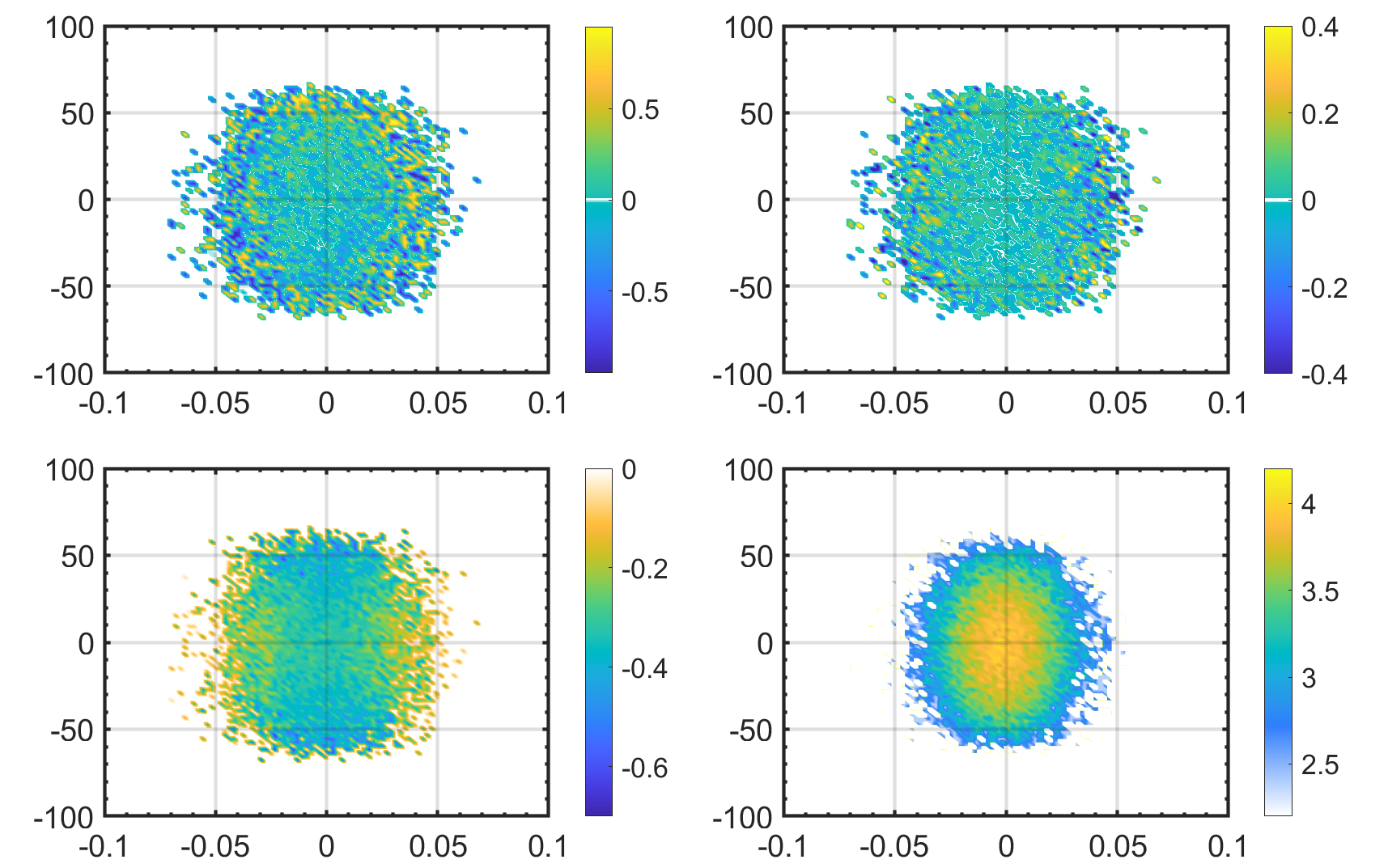}\\
  \begin{picture}(300,20) 
   \put(22,160){(a)}
    \put(-4,124){\rotatebox{90}{$\theta_x$(mrad) }}
    \put(44,94){$\theta_y$(mrad) }
 \put(142,160){(b)}
 \put(122,124){\rotatebox{90}{$\theta_x$(mrad) }}
 \put(165,94){$\theta_y$(mrad) }
 \put(22,83){(c)}
    \put(-4,42){\rotatebox{90}{$\theta_x$(mrad)}}
    \put(44,11){$\theta_y$(mrad) }
    \put(142,83){(d)}
    \put(120,42){\rotatebox{90}{$\theta_x$(mrad)}}
    \put(165,11){$\theta_y$(mrad) }
 \end{picture}\\
    \caption{ The angular distribution of the polarization of positrons with $\varepsilon_+>1$~GeV: 
    (a) for $\zeta_x$, (b) for  $\zeta_y$, (c) for $\zeta_z$ . (d) The angular distribution of the number density  $d^2N/d\theta_xd\theta_y$ for positrons with $\varepsilon_+>1$~GeV.}
     \label{Fig.pos_noradiation}
\end{figure}
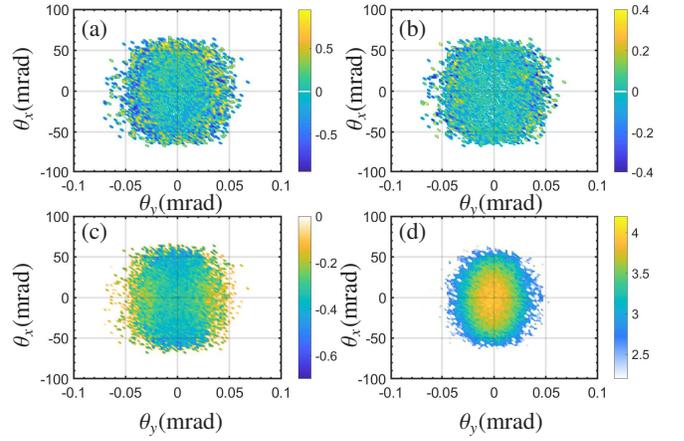
\begin{figure}[b]
 \includegraphics[width=0.48\textwidth]{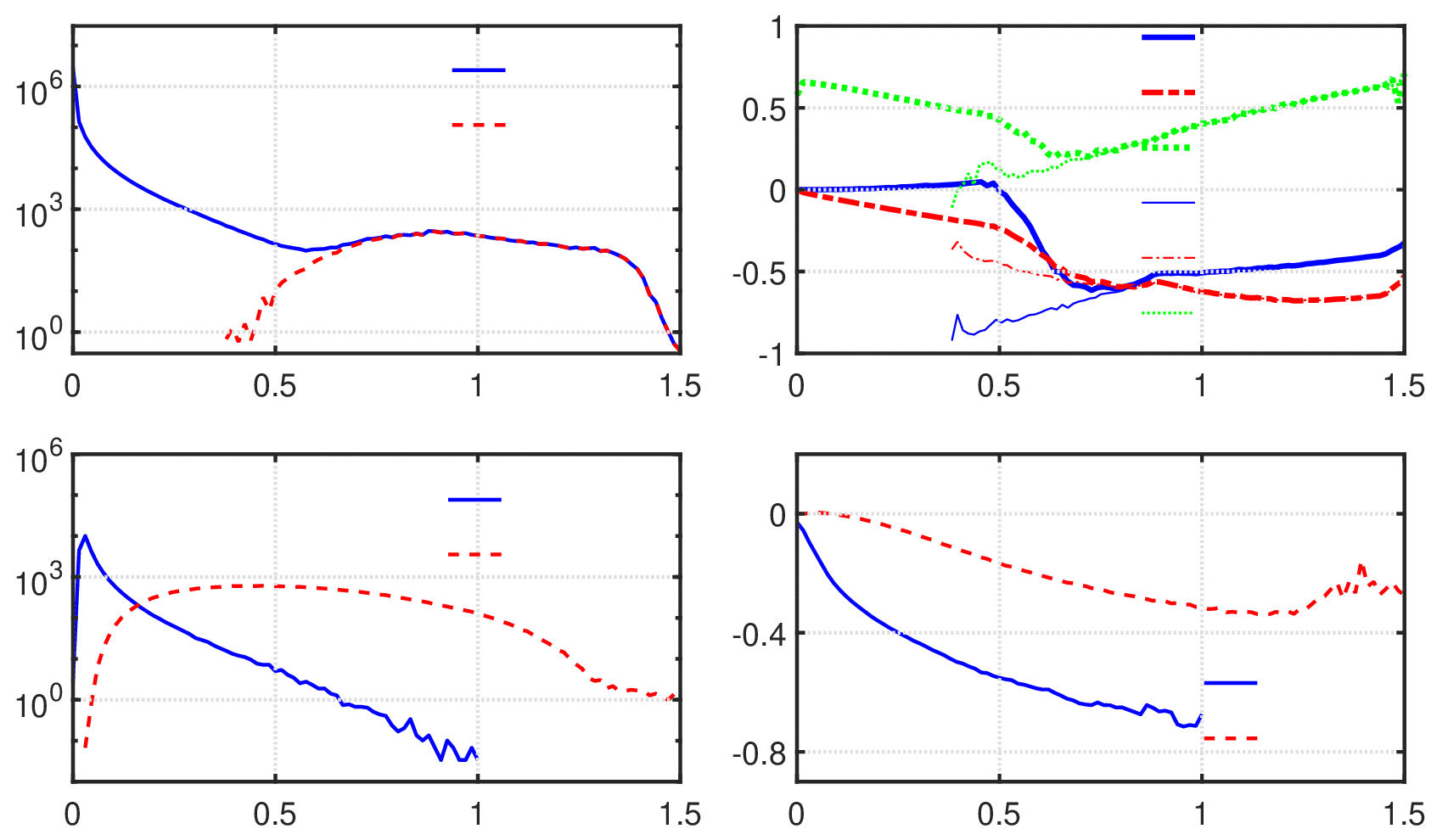}\\
  \begin{picture}(300,20)
     \put(17,150){(a)}
     \put(60,90){$\omega$}
     \put(-7,110){\rotatebox{90}{$mdN_\gamma/d\omega$}}
    \put(89,149){\fontsize{7pt}{\baselineskip}\selectfont{with rad}}
    \put(89,140){\fontsize{7pt}{\baselineskip}\selectfont{w/o rad}}    
   
    \put(140,150){(b)}
    \put(185,90){$\omega$}
    \put(120,128) {\rotatebox{90}{$\xi$}}
    \put(205,154){\fontsize{4.7pt}{\baselineskip}\selectfont{$\xi_1$}}
    \put(205,146){\fontsize{4.7pt}{\baselineskip}\selectfont{$\xi_2$}}
    \put(205,136){\fontsize{4.7pt}{\baselineskip}\selectfont{$\xi_3$}}
    \put(205,127){\fontsize{4.7pt}{\baselineskip}\selectfont{$\xi_1$ w/o rad }}
    \put(205,119){\fontsize{4.7pt}{\baselineskip}\selectfont{$\xi_2$ w/o rad }}
    \put(205,108){\fontsize{4.7pt}{\baselineskip}\selectfont{$\xi_3$ w/o rad }}
    
    \put(17,78){(c)}
    \put(60,15){$\varepsilon_+$}
    \put(-7,39){\rotatebox{90}{$mdN_{e^+}/d\varepsilon_+$}}
    \put(88,77){\fontsize{7pt}{\baselineskip}\selectfont{with rad}}
    \put(88,67){\fontsize{7pt}{\baselineskip}\selectfont{w/o rad}}
     
    \put(140,78){(d)}
    \put(185,13){$\varepsilon_+$}
    \put(118,57){\rotatebox{90}{$P_\parallel$}}
    \put(214,46){\fontsize{7pt}{\baselineskip}\selectfont{with rad}}
    \put(214,36){\fontsize{7pt}{\baselineskip}\selectfont{w/o rad}}
    \end{picture}\\
  \caption{ (a) Photon spectrum $mdN_\gamma/d\omega$ with (blue solid line) and without (red dashed line) pair radiation. (b) The distribution of photon polarization versus photon energy $\omega$ (GeV) with (thick lines) and without (thin lines) radiation from pairs  for: $\xi_1$ (blue solid line), $\xi_2$ (red dot-dashed line), and $\xi_3$ (green dotted line).  (c)  Positron spectrum $mdN_{e^+}/d\varepsilon_+$ (d) Longitudinal polarization of positrons $P_\parallel$ vs $\varepsilon_+$ (GeV) taking into account pair radiation (blue solid line), or neglecting it (red dashed line).  The laser pulse duration $\tau_p=50$ fs.}
  \label{Fig.phoposunrad}
\end{figure}
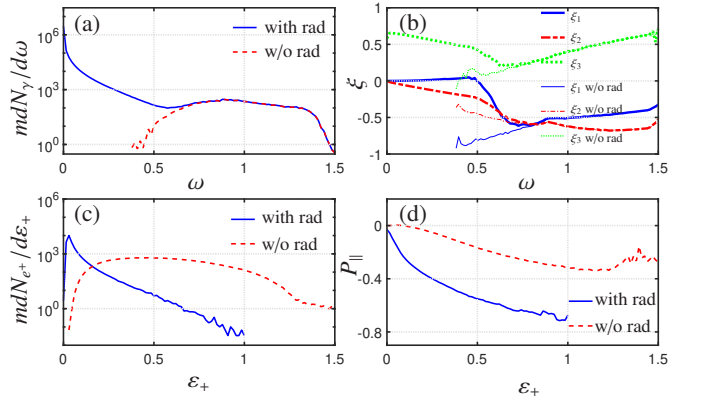

\subsection{Impact of secondary photon emissions}

The impact of secondary emissions on the photonic signal are shown in Fig. \ref {Fig.phoposunrad} (a) and (b). The emissions of pairs extend the spectrum to the low-energy region [Fig. \ref {Fig.phoposunrad}(a)] and significantly affect the average polarization around 0.5~GeV [Fig. \ref {Fig.phoposunrad} (b)]. The emitted photons are linearly polarized with $\overline{\xi}_3\approx 59\%$, see Fig. \ref {Fig.phoposunrad} (a).
Fortunately, the polarization and spectrum in the high energy region are not affected by the radiation of pairs. If the gamma photons with energy higher than 0.75~GeV are post-selected, a clean signal of vacuum polarization can be obtained. Otherwise, the low-energy photons will overwhelm the VP photonic signal.

\begin{figure}[b]
 \includegraphics[width=0.48\textwidth]{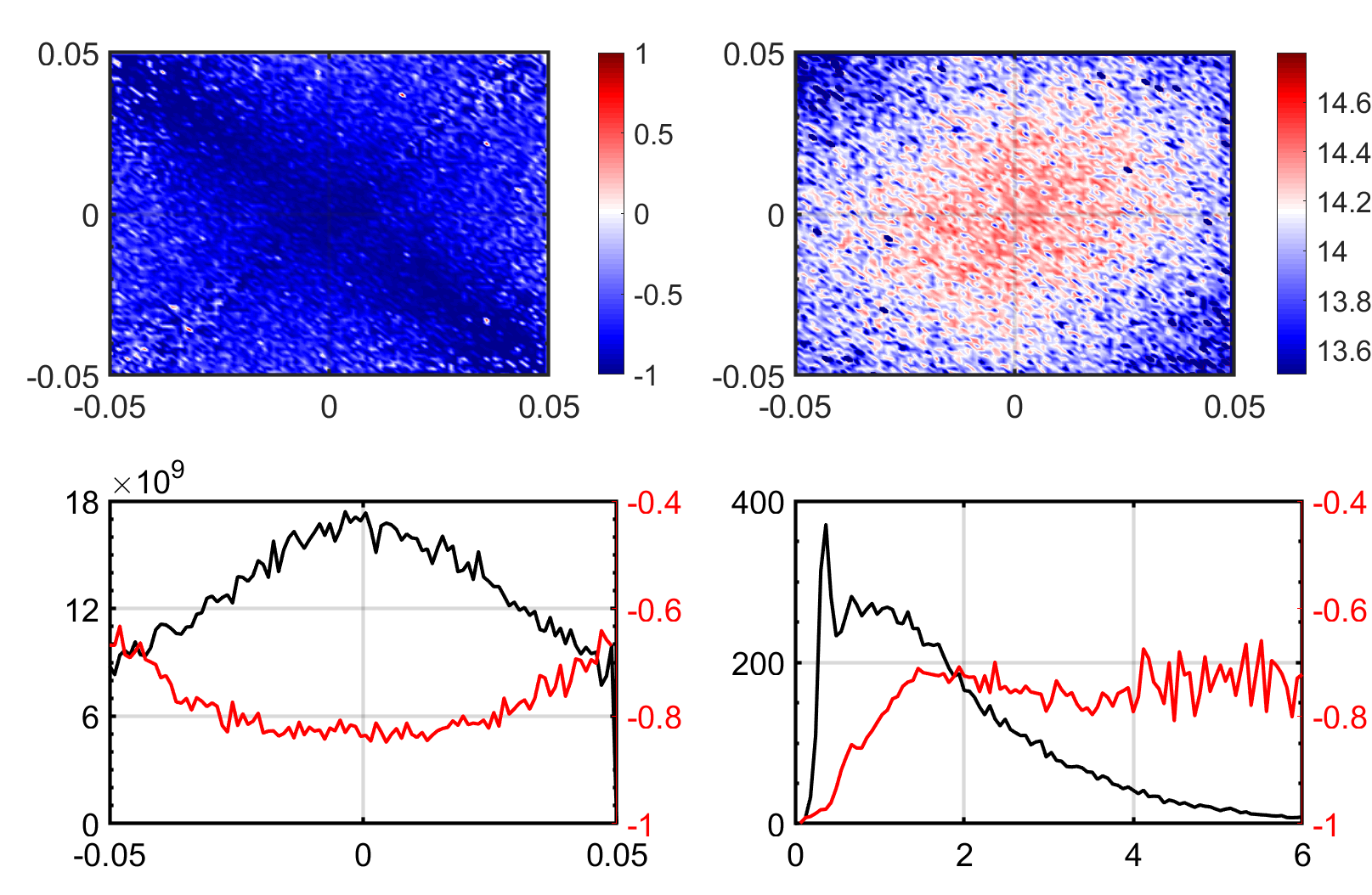}\\
  \begin{picture}(300,20)    
     \put(22,164){\color{white}(a)}
    \put(58,99){\small $\theta_y$}
    \put(2,143) {\rotatebox{90}{\small $ \theta_x$}}
    
	\put(148,164){\color{white}(b)}
	\put(179,99){\small $\theta_y$}
    \put(127,143) {\rotatebox{90}{\small $ \theta_x$}}
    
    \put(23,83){(c)}
    \put(63,15){\small $\theta_x$}
    \put(2,50){\rotatebox{90}{\small $dN_\gamma/d\theta_x$}}
     
	\put(219,83){(d)}
	\put(178,15){\small $\omega$ (GeV)}
	\put(123,47){\rotatebox{90}{\small m$dN_\gamma/d\omega$}}
    \put(242,66){\rotatebox{270}{\small\color{black}$\bar{\xi_1}$}}
	\end{picture}\\
   \caption{ (a) Angular distribution of $\gamma$ photon density $\text{log}_{10}d^2N/d\theta_x/d\theta_y$ $ (\text{mrad}^{-2})$ and (b) polarization $\xi_1$ vs $\theta_x$ (mrad) and $\theta_y$ (mrad) obtained with parameters of LUXE:  $a_0=3\sqrt{2}$, $\varepsilon_i=17.5$ GeV, pulse duration 30 fs. (c) The angular distribution of $\gamma$ photon density $dN_\gamma/d\theta_x$ $ (\text{mrad}^{-1})$ (black solid line) and polarization $\xi_1$ (red solid line) vs $\theta_x$.  (d) The energy distribution of $\gamma$ photon density m$dN_\gamma/d\omega$ $ (\text{GeV}^{-1})$ (black solid line) and polarization $\xi_1$ (red solid line) vs $\omega$ $(\text{GeV})$.}
   \label{LUXE_pho}
\end{figure}
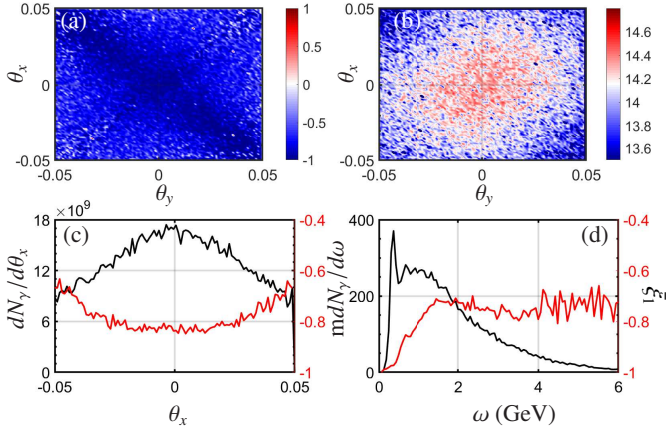

 \begin{figure}[b]
 \includegraphics[width=0.48\textwidth]{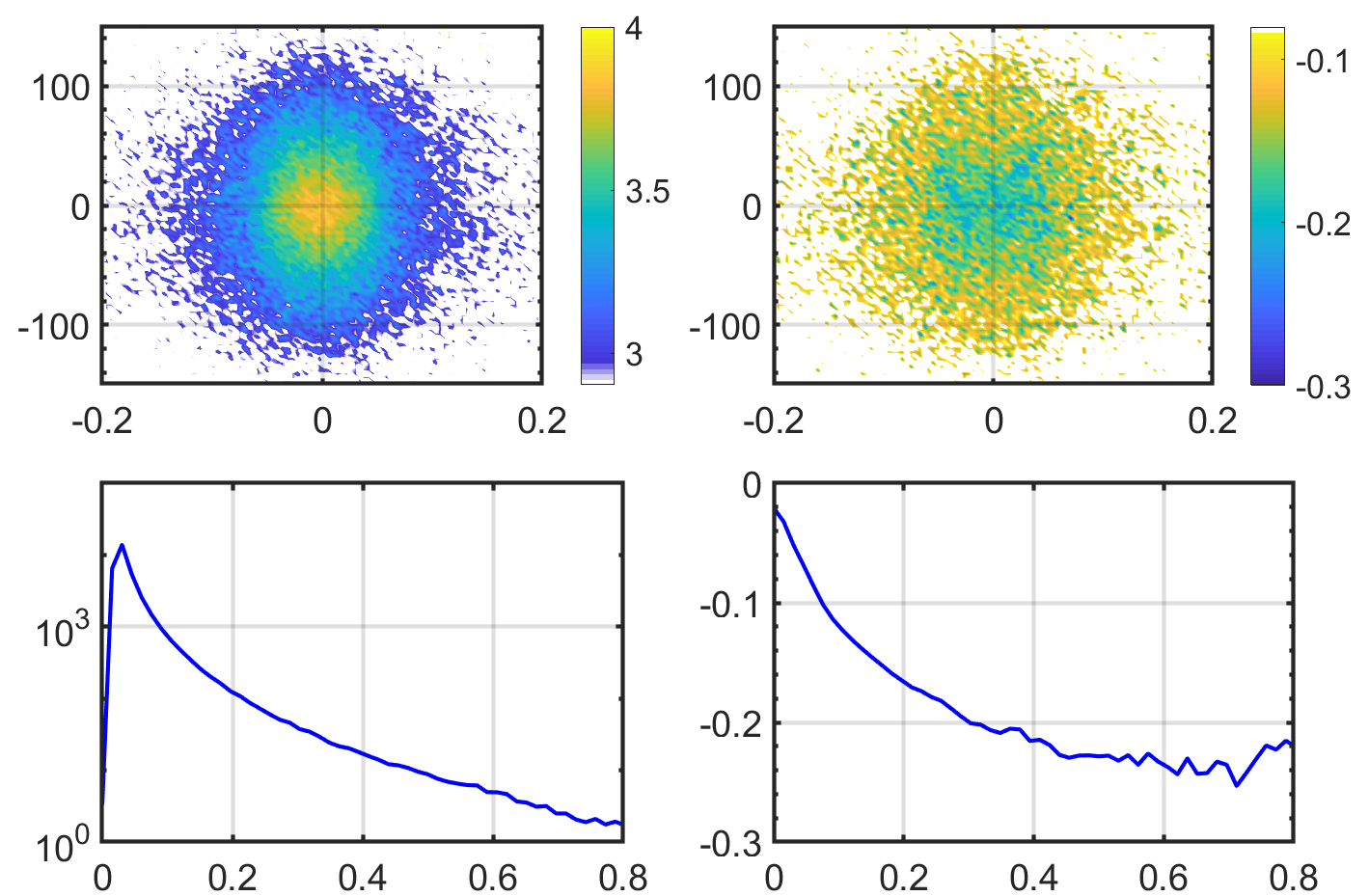}\\
  \begin{picture}(300,20)    
    \put(22,169){(a)}
    \put(-4,128){\rotatebox{90}{$\theta_x$(mrad) }}
    \put(44,99){$\theta_y$(mrad) }
	\put(142,169){(b)}
	\put(122,128){\rotatebox{90}{$\theta_x$(mrad) }}
	\put(165,99){$\theta_y$(mrad) }
	\put(101,87){(c)}
    \put(-4,42){\rotatebox{90}{$mdN_{e^+}/d\varepsilon_+$}}
    \put(50,11){$\varepsilon_+$ (GeV) }
    \put(220,87){(d)}
    \put(120,58){\rotatebox{90}{$P_\parallel$}}
    \put(170,11){$\varepsilon_+$ (GeV) }
	\end{picture}\\
 \caption{The positron angular distribution produced by photons of Fig .\ref{LUXE_pho}: (a) for the number density $d^2N_{e^+}/d\theta_xd\theta_y$ (mrad$^{-2}$), (b) for the longitudinal polarization $P_\parallel$, when $\theta_{x,y}$  are in mrad. Positron number density $mdN_{e^+}/d\varepsilon_+$  (c), and the longitudinal polarization (d) vs positron energy $\varepsilon_+$ (GeV). $N_{\gamma0}=8\times 10^5$.}
 \label{LUXE_pos}
\end{figure}
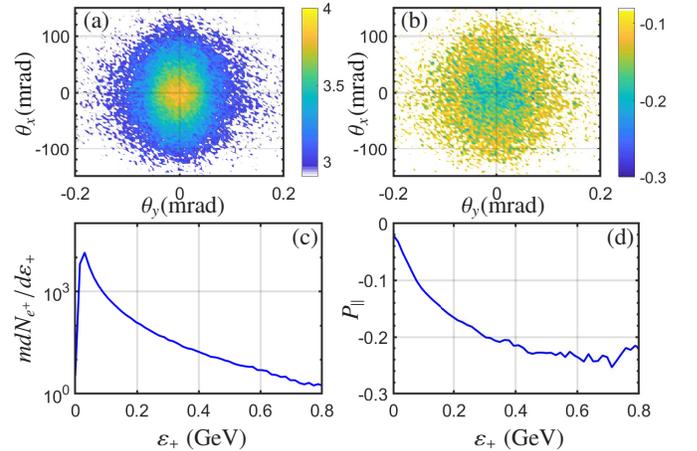
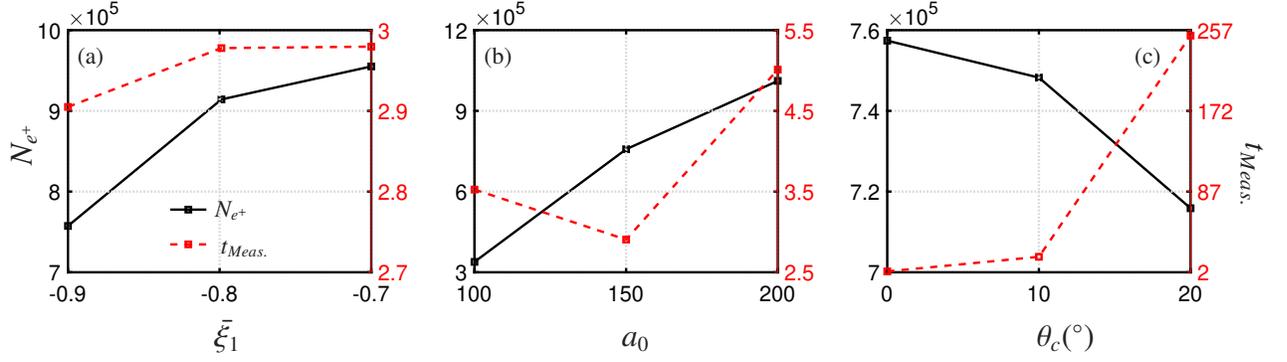
\begin{figure*}
   \includegraphics[width=0.9\textwidth]{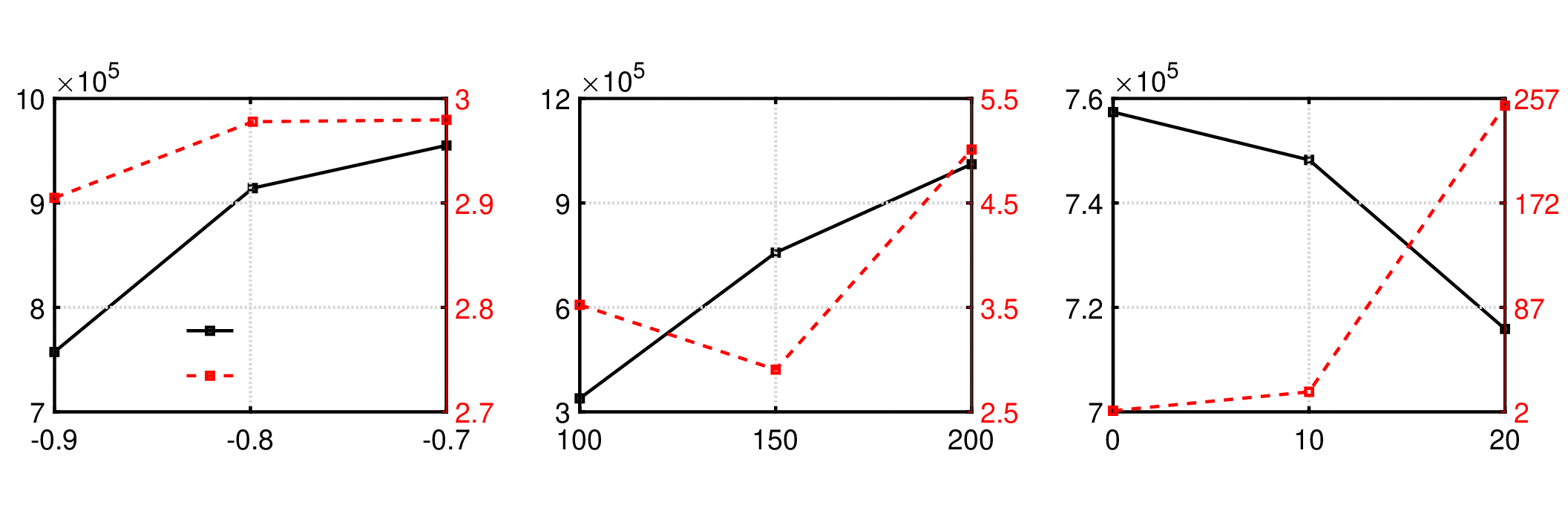}
  \begin{picture}(300,20)  
    \put(-85,91){\rotatebox{90}{\large $N_{e^+}$}}  
    \put(380,100){\rotatebox{270}{\large $t_{Meas.}$}}
    \put(-60,130){(a)}
    \put(-8,72){$N_{e^+}$}
    \put(-8,58){ $t_{Meas.}$}
    \put(-8,23){\large $\bar{\xi_1}$}
 \put(94,130){(b)}
    \put(146,23){\large $a_0$}
    \put(340,130){(c)}
    \put(303,23){\large $\theta_c(^\circ)$}
 \end{picture}\\
  \caption{The scaling laws of positron yield $N_e^+$ (black solid line) and meaurement time $t_{\text{Meas.}}$ in unit of hour (red dashed line)  versus  (a) the polarization of the initial gamma photons $\bar{\xi_1}$, (b) laser intensity $a_0$, and (c) the collision angle $\theta_c$ between laser and the $\gamma$-ray beam.}
    \label{scaling_law}
\end{figure*}

The impact of secondary emissions on the fermionic signal is shown in Fig. \ref {Fig.phoposunrad} (c) and (d). The radiation of pairs results in a redistribution of positron energy, see Fig. \ref {Fig.phoposunrad} (c) and (d). Without radiation, the positrons exhibit a wide energy distribution, extending up to 1.5 GeV. However, when radiation reaction is included, the energy distribution of the positrons peaks at 30 MeV. Moreover, the secondary emission alters the distribution of polarization. The maximum polarization increases from $34\%$ to $71\%$ when taking into account secondary emissions. This can be explained as follows. Without radiation, positrons with different polarizations are mixed, resulting in a relatively low average polarization [see Fig. \ref {Fig.pos_noradiation} (c)]. 
The positrons at larger $\theta_y$ have smaller longitudinal polarization [Fig.~\ref {Fig.pos_noradiation}~(c)] but higher transverse polarization [Figs. \ref {Fig.pos_noradiation} (a) and (b)], and vice versa at smaller $\theta_y$.
While positrons with polarization levels of up to $70\%$ already exist, they are overwhelmed by the large number of positrons with lower polarization.
When radiation is taken into account, the positrons with different polarization are separated due to the spin-dependent radiation probability, i.e. $dW_{rad}\propto dW^0 -\frac{\omega}{\varepsilon_+}\zeta_y \text{K}_{\frac{1}{3}}(z_q)$ with $ dW^0$ being the unpolarized radiation probability. Specifically, positrons with large negative $\zeta_y$ (and correspondingly small $\zeta_z$) undergo more dramatic radiation reactions and are therefore more significantly red-shifted. As the components with low polarization are reduced, $\overline{\zeta}_z$ at the high-energy end of the spectrum increases [Fig. \ref{Fig.phoposunrad} (d)]. 

Even through the maximum of polarization increases, the average polarization decreases from $17\%$ to $13\%$. This is confirmed by the following equation describing the evolution of the average longitudinal polarization \cite{li2022helicity}
\begin{align*}\label{sparallel}
\frac{dP_{\parallel}}{dt}&=-\frac{e}{m}\boldsymbol{P}_{\perp}\cdot\left[\left(\frac{g}{2}-1\right)\boldsymbol{\beta\times B}+\left(\frac{g\beta}{2}-\frac{1}{\beta}\right)\boldsymbol{E}\right]\\&-\frac{\alpha m}{\sqrt{3}\pi\gamma}P_{\parallel}\int_{0}^{\infty}\frac{u^{2}du}{\left(1+u\right)^{3}\int_{z_{q}}^{\infty}dx\textrm{K}_{\frac{1}{3}}\left(x\right)}\\&\approx-\frac{\alpha m}{\sqrt{3}\pi\gamma}P_{\parallel}\int_{0}^{\infty}\frac{u^{2}du}{\left(1+u\right)^{3}\int_{z_{p}}^{\infty}dx\textrm{K}_{\frac{1}{3}}\left(x\right)},\numberthis
\end{align*}
where $u=\omega'/\varepsilon'$, and the last term is due to radiation. The approximation in Eq. (\ref{sparallel}) is justified because $\boldsymbol{P}_{\perp}\cdot\boldsymbol{E}=0$ for radiative polarization in linearly polarized laser fields. According to Eq. (\ref{sparallel}), the longitudinal polarization $|dP_{\parallel}|$ decreases due to radiation.

\subsection{Impact of the initial gamma beam parameters}

\subsubsection{Initial gamma photons energy} 

We have employed the relatively low-energy electrons (LEPS2 beamline at SPring-8) because the photon energy obtained by perfect backscattering of 8.4 GeV electron is $\omega=\left(1+\beta\right)\varepsilon\omega_{0}/(\varepsilon-\varepsilon\beta+2\omega_{0})\approx 1.13$ GeV, which is within the optimal energy range for enhancing signal of vacuum polarization.

Can a better result be obtained with a more advanced electron source, e.g., LUXE? 
With the high-energy electrons at LUXE (17.5 GeV), the interaction enters the nonlinear nonperturbative regime, where the photon density and energy increase, however, at the expense of a decrease in polarization. The production rate of photons increases from $N_\gamma\approx0.5N_{e_0^-}$ to  $N_\gamma\approx0.66N_{e_0^-}$ [Figs. \ref{LUXE_pho} (b) and (c)], while the average polarization of positrons decreases from $|\xi_1|=0.91$ to  $\xi_1=0.78$ [Fig. \ref{LUXE_pho} (a)]. Meanwhile, the photon spectrum undergoes broadening to 6 GeV [Fig. \ref{LUXE_pho} (d)]. 
Hence, the photons obtained under the parameters of 
the LUXE project fall outside the optimal range for conducting vacuum polarization measurements. For instance, 
with the probe gamma photons obtained with 17.5 GeV electrons, the positrons number increases from $N_{e^+}=3.8\times 10^7$ to $8.2\times 10^7$ for initial electrons $N_{e^-_0}=10^8$ [Figs. \ref{LUXE_pos} (a) and (c)], while the longitudinal polarization of produced positrons decreases from 13\%  to 6.4\%  [Figs. \ref{LUXE_pos} (b) and (d)]. The substantial decreases in polarization leads to a longer measurement time, $t_\text{meas.}$=6.7 hours. 

\subsubsection{Initial gamma photon polarization} 

The variability in the collection angle of photons could introduce uncertainty to the polarization of the gamma-ray beam. As the collision angle of the gamma-ray beam increases from $\Delta\theta_{\text{max}}=0.05$ mrad to 0.1 mrad, the photon yield increases while the average polarization decreases from $|\xi_1|=0.9$ to 0.7. This decline in photon polarization results in an extended measurement time for vacuum polarization. However, this is counterbalanced by the enhanced positron yield. Consequently, the measurement time increases slightly from 2.9 to 3 hours as the polarization degree decreases, see Fig. \ref{scaling_law} (a).

\begin{figure} 
 \includegraphics[width=0.48\textwidth]{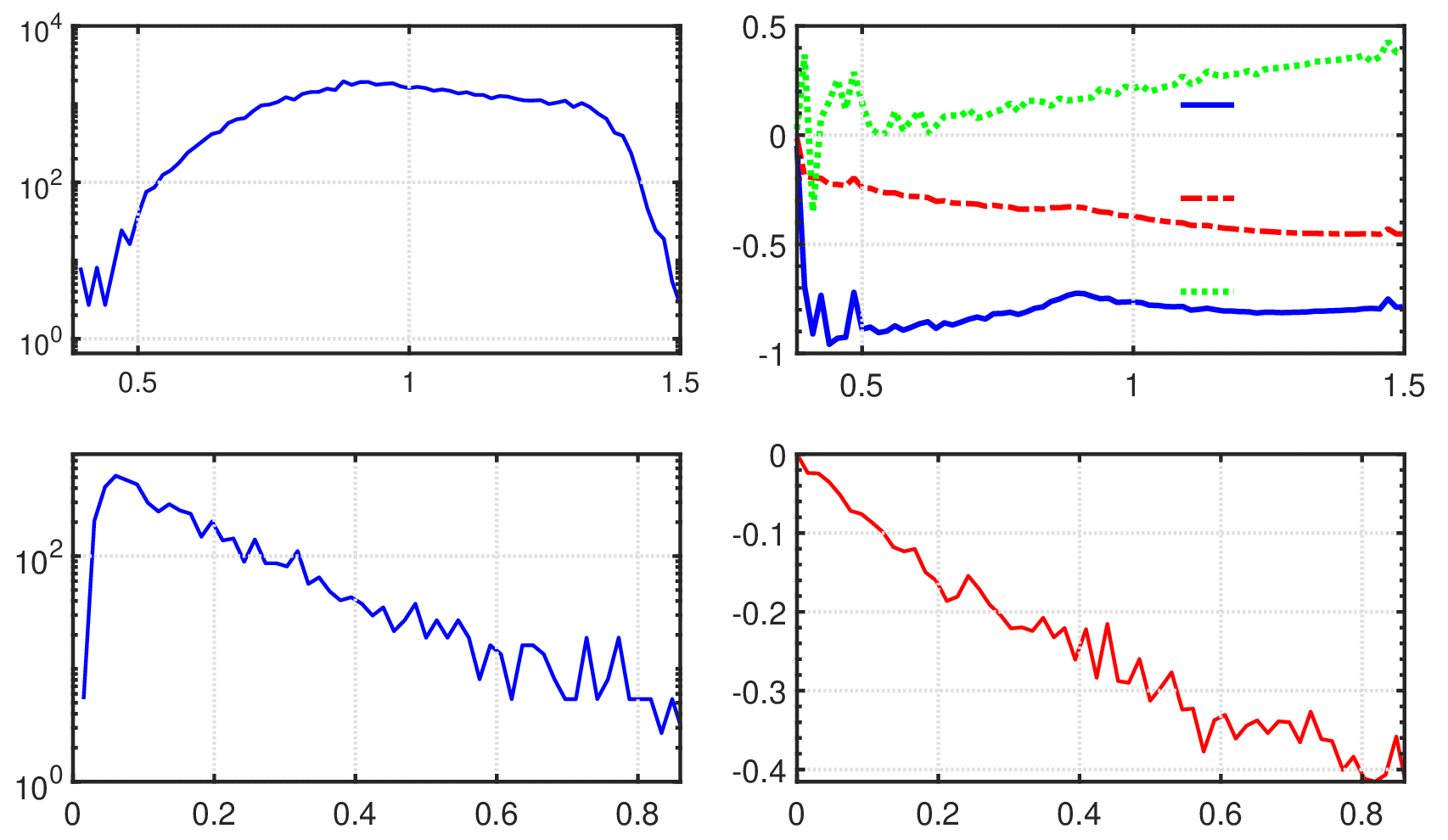}\\
  \begin{picture}(300,20)
      \put(17,150){(a)}
     \put(60,90){$\omega$}
     \put(-7,110){\rotatebox{90}{$mdN_\gamma/d\omega$}}
   
    \put(142,150){(b)}
    \put(185,90){$\omega$}
    \put(120,128) {\rotatebox{90}{$\xi$}}
    \put(213,144){\fontsize{6pt}{\baselineskip}\selectfont{$\xi_1$}}
    \put(213,128){\fontsize{6pt}{\baselineskip}\selectfont{$\xi_2$}}
    \put(213,113){\fontsize{6pt}{\baselineskip}\selectfont{$\xi_3$}}
    
    \put(101,78){(c)}
    \put(60,15){$\varepsilon_+$}
    \put(-7,39){\rotatebox{90}{$mdN_{e^+}/d\varepsilon_+$}}

    \put(222,78){(d)}
    \put(185,13){$\varepsilon_+$}
    \put(118,55){\rotatebox{90}{$P_\parallel$}}

    \end{picture}\\
    \caption{ Photon density $mdN_\gamma/d\omega$ (a) and polarization distribution (b) within 0.05 mrad vs photon energy $\omega$ (GeV) for: $\xi_1$ (blue solid line),  $\xi_2$ (red dot-dashed line), $\xi_3$ (green dotted line). Positron density $mdN_{e^+}/d\varepsilon_+$ (c) and the distribution of longitudinal polarization (d) within 10 mrad vs positron energy $\varepsilon_+$ (GeV). The laser pulse duration $\tau$ = 25 fs.}
    \label{Fig.phopos25fs}
\end{figure}
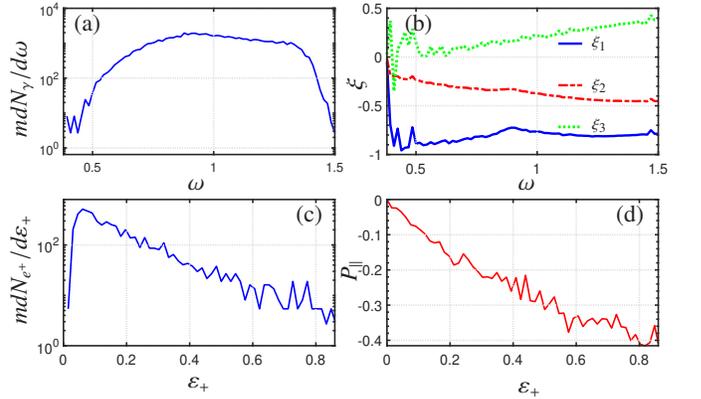
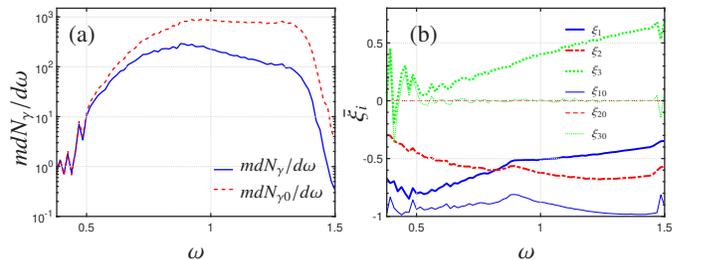
\begin{figure}  
	\includegraphics[width=0.48\textwidth]{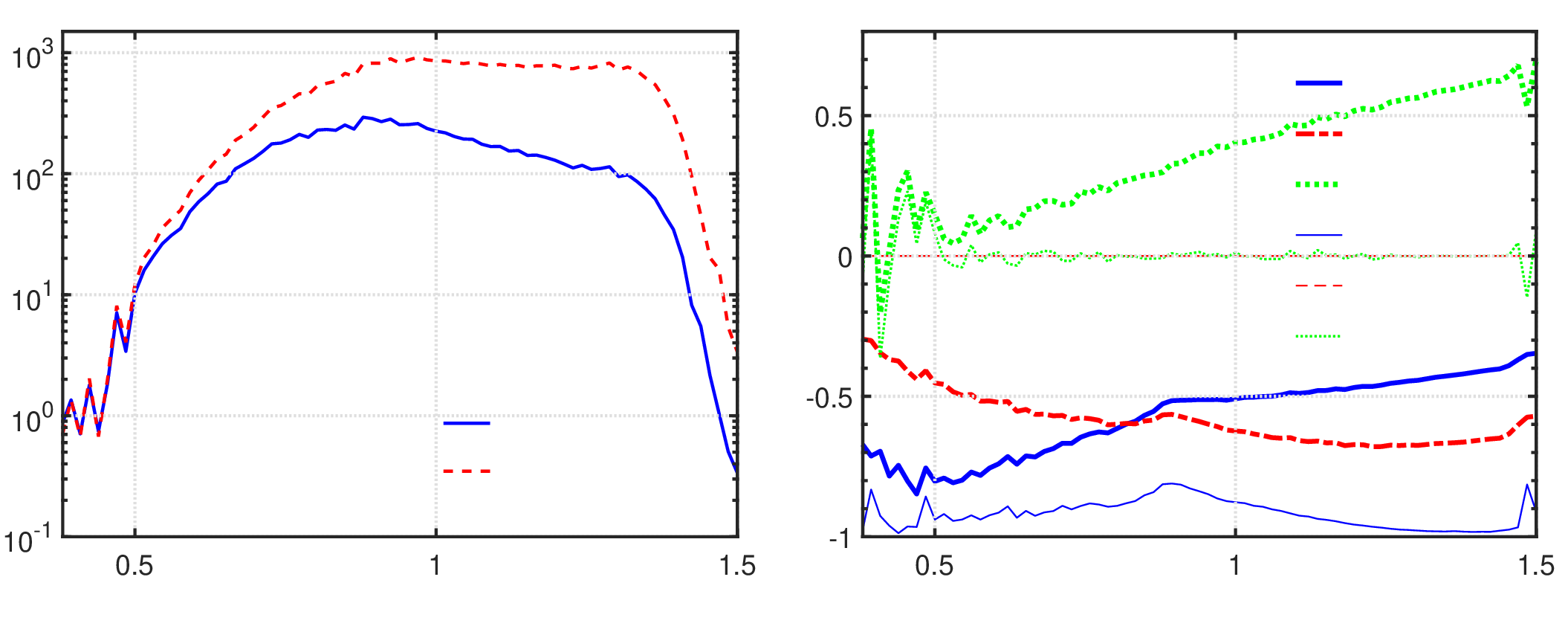}\\
    \begin{picture}(300,20)    
     \put(15,102){(a)}
     \put(60,20){$\omega$}
     \put(-7,55){\rotatebox{90}{$mdN_\gamma/d\omega$}}
     \put(80,53){\fontsize{7pt}{\baselineskip}\selectfont{$mdN_{\gamma}/d\omega$}}
     \put(80,43){\fontsize{7pt}{\baselineskip}\selectfont{$mdN_{\gamma0}/d\omega$}}   
          
    \put(144,102){(b)}
    \put(185,20){$\omega$}
    \put(119,72){\rotatebox{90}{$\bar{\xi_i}$}}
     \put(213,105){\fontsize{4.7pt}{\baselineskip}\selectfont{$\xi_1$}} 
     \put(213,97){\fontsize{4.7pt}{\baselineskip}\selectfont{$\xi_2$}}  
     \put(213,90){\fontsize{4.7pt}{\baselineskip}\selectfont{$\xi_3$}}  
     \put(213,82){\fontsize{4.7pt}{\baselineskip}\selectfont{$\xi_{10}$}}  
     \put(213,74){\fontsize{4.7pt}{\baselineskip}\selectfont{$\xi_{20}$}}  
     \put(213,66){\fontsize{4.7pt}{\baselineskip}\selectfont{$\xi_{30}$}}   

    \end{picture}\\
    \caption{ (a) Density of probe gamma photons before (red dashed line)  and after (blue solid line) interacting with the laser.  (b) Polarization of probe gamma photons before (thin line) and after (thick line) interacting with the laser. The laser pulse duration is $\tau_p=50$ fs. }
    \label{Fig.pho1d}
\end{figure}

\subsection{Impact of the laser parameters}

\subsubsection{Pulse duration} 

The duration of the laser pulse controls the conversion of the circularly polarized gamma-photons into the longitudinally polarized positrons, and determines the balance between the photonic and fermionic signals of VP. 


The effect of VP is less significant in a shorter laser pulse with a pulse duration of $\tau_p=25$~fs, compared to the 50 fs case discussed so far, cf. Fig.~\ref{Fig.phopos25fs} with Fig.~\ref{Fig.pho1d}. However, the number of survived outgoing photons is larger. Thus, in the considered scenario, half of the probe photons decay into pairs, while the other half survive without undergoing pair production. For an initial count of $N_{e^-}^0=1\times 10^8$ electrons, we are left with $N_\gamma=2.5\times 10^7$ probe photons available for measuring vacuum polarization. Even though the photon yield is higher compared to the $\tau_p=50$~fs case [Fig. \ref{Fig.phopos25fs} (a) cf.  Fig.~\ref{Fig.pho1d}(a)], the variation in polarization induced by vacuum polarization is smaller, due to the reduced interaction length [Fig. \ref{Fig.phopos25fs} (b) cf. Fig.~\ref{Fig.pho1d}(b)]. The average photon polarization in the small-angle region ($\theta<0.05$mrad) becomes $\xi=(78\%,37\%,21\%)$. In this case, employing the polarimetry method outlined in Sec. IV B, a single-shot measurement could achieve a confidence level of $3\sigma$ for vacuum birefringence and $6\sigma$ for vacuum dichroism.

\begin{figure} 
 \includegraphics[width=0.48\textwidth]{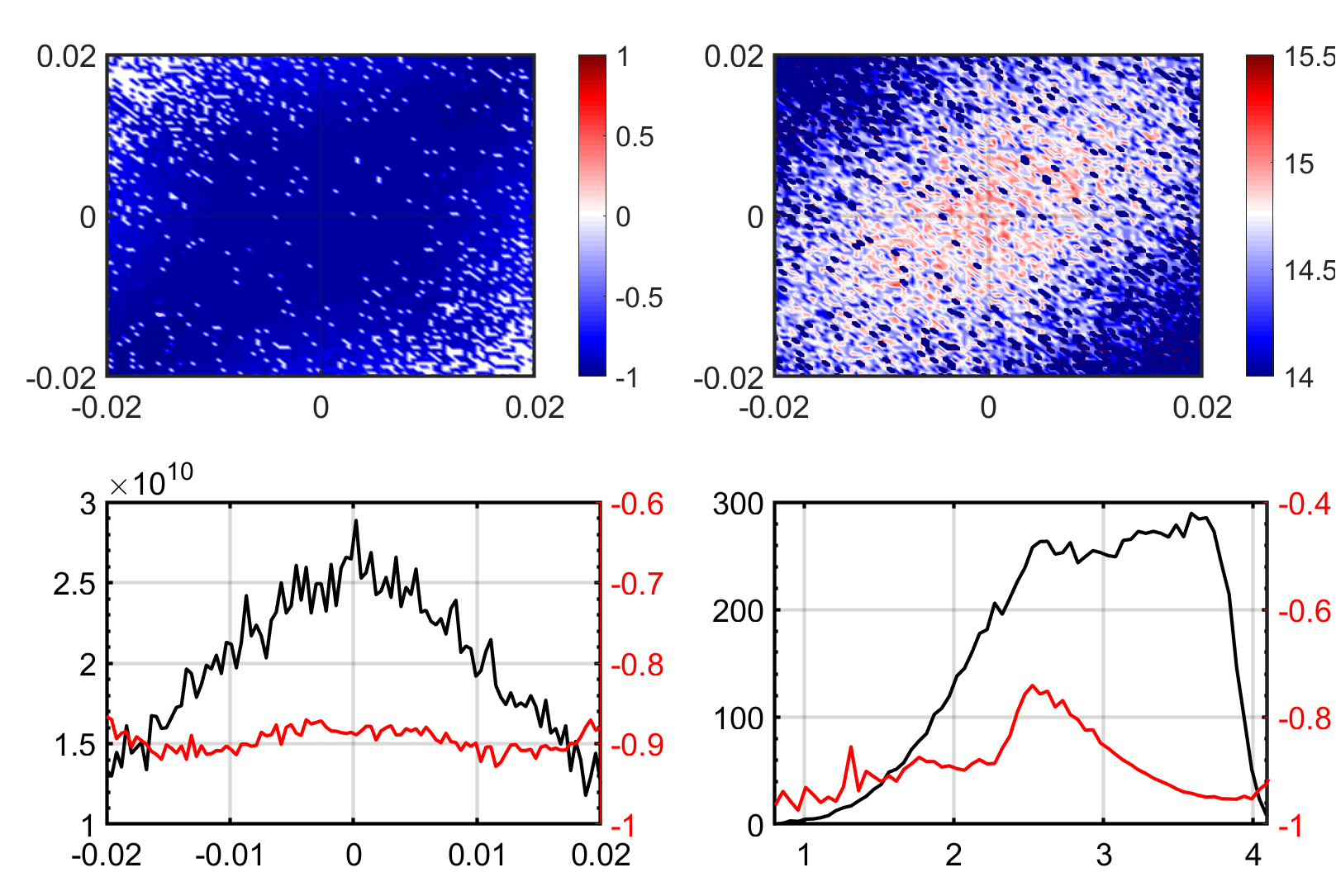}\\
  \begin{picture}(300,20)    
    \put(22,164){\color{white}(a)}
    \put(58,99){\small $\theta_y$}
    \put(2,143) {\rotatebox{90}{\small $ \theta_x$}}
    
	\put(148,164){\color{white}(b)}
	\put(179,99){\small $\theta_y$}
    \put(127,143) {\rotatebox{90}{\small $ \theta_x$}}
    
    \put(23,83){(c)}
    \put(63,15){\small $\theta_x$}
    \put(2,50){\rotatebox{90}{\small $dN_\gamma/d\theta_x$}}
     
	\put(148,83){(d)}
	\put(178,15){\small $\omega$ (GeV)}
	\put(123,47){\rotatebox{90}{\small m$dN_\gamma/d\omega$}}
    \put(242,66){\rotatebox{270}{\small\color{black}$\bar{\xi_1}$}}
	\end{picture}\\
 \caption{(a) Angular distribution of $\gamma$ photon density $\text{log}_{10}d^2N/d\theta_x/d\theta_y$ $ (\text{mrad}^{-2})$ and (b) polarization $\xi_1$ vs $\theta_x$ (mrad) and $\theta_y$ (mrad). (c) The angular distribution of $\gamma$ photon density $dN_\gamma/d\theta_x$ $ (\text{mrad}^{-1})$ (black solid line) and polarization $\xi_1$ (red solid line) vs $\theta_x$.  (d) The energy distribution of $\gamma$ photon density m$dN_\gamma/d\omega$ $ (\text{GeV}^{-1})$ (black solid line) and polarization $\xi_1$ (red solid line) vs $\omega$ $(\text{GeV})$. The initial electron energy is 15GeV.}
 \label{1pw_pho}
\end{figure}

 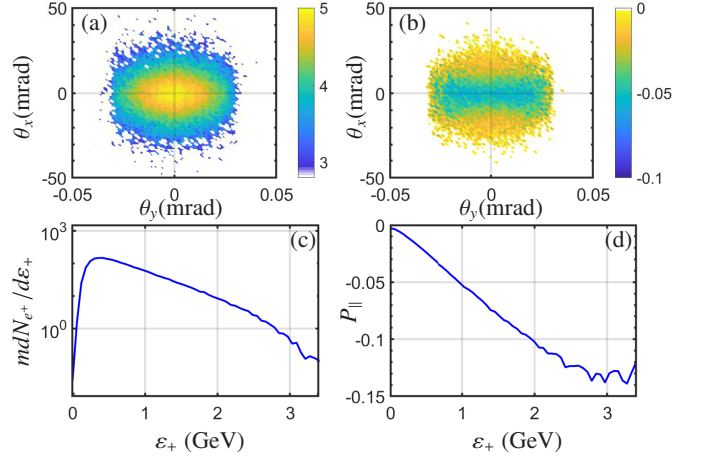
\begin{figure}
 \includegraphics[width=0.48\textwidth]{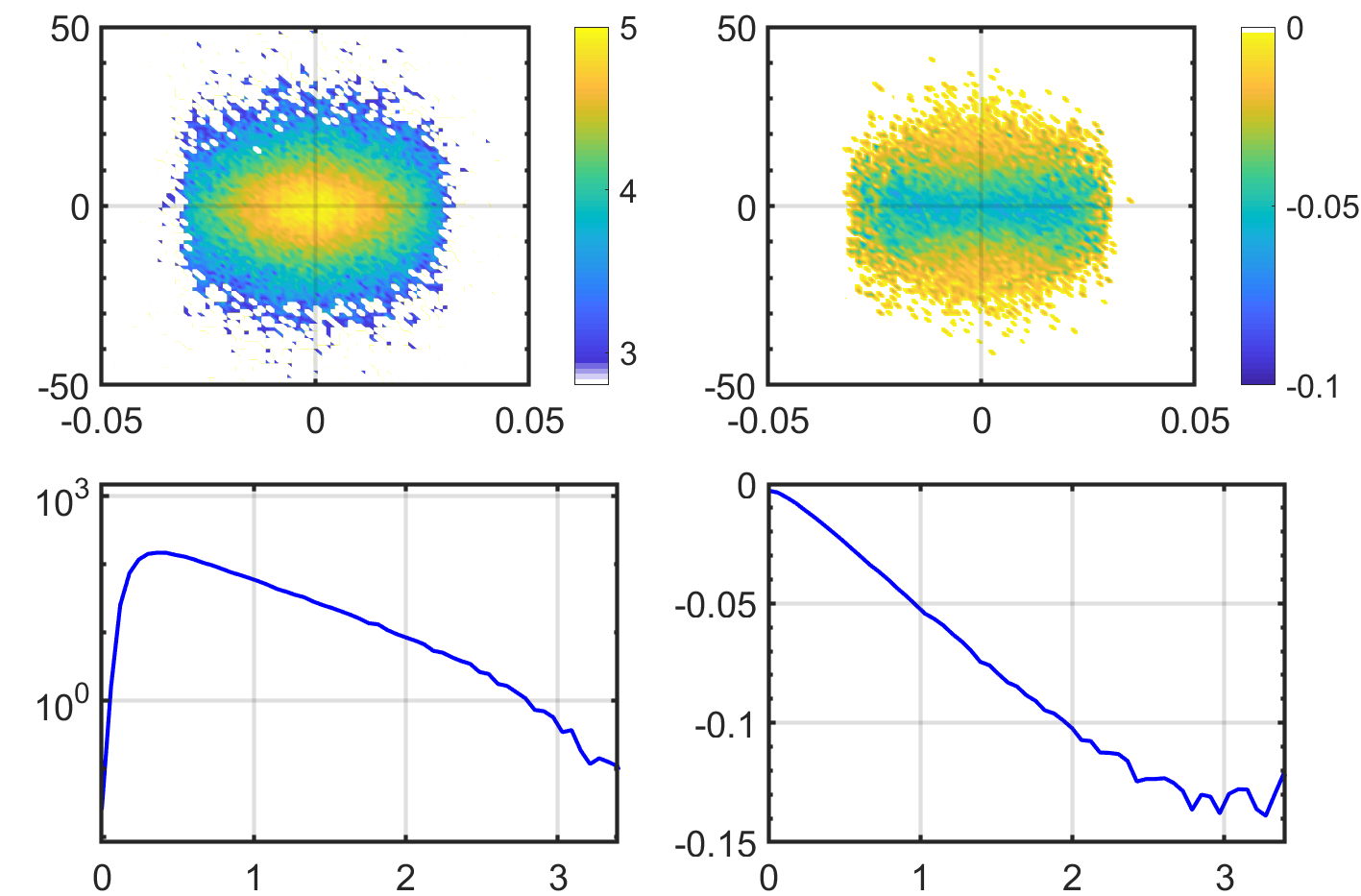}\\
  \begin{picture}(300,20)    
   \put(22,169){(a)}
    \put(-4,128){\rotatebox{90}{$\theta_x$(mrad) }}
    \put(44,99){$\theta_y$(mrad) }
	\put(142,169){(b)}
	\put(122,128){\rotatebox{90}{$\theta_x$(mrad) }}
	\put(165,99){$\theta_y$(mrad) }
	\put(101,87){(c)}
    \put(-4,42){\rotatebox{90}{$mdN_{e^+}/d\varepsilon_+$}}
    \put(50,11){$\varepsilon_+$ (GeV) }
    \put(220,87){(d)}
    \put(120,58){\rotatebox{90}{$P_\parallel$}}
    \put(170,11){$\varepsilon_+$ (GeV) }
	\end{picture}\\
  \caption{ (Top row) The positron angular distribution: (a) for the number density $d^2N_{e^+}/d\theta_xd\theta_y$ (mrad$^{-2}$), (b) for the longitudinal polarization $P_\parallel$, when $\theta_{x,y}$  are in mrad.  (Bottom row) Positron number density $mdN_{e^+}/d\varepsilon_+$  (c), and the longitudinal polarization (d) vs positron energy $\varepsilon_+$ (GeV). Laser intensity $a_0 = 50$ and pulse duration $t_p = 10T$.}
  \label{1pw_pos}
\end{figure}

Meanwhile, as the pulse duration decreases from 50 fs to 25 fs, the positron yield reduces from $N_{e^+e^-}=7.5\times 10^7$ to $N_{e^+e^-}=2.5\times 10^7$ [Fig. \ref{Fig.phopos25fs} (c)]. The fermionic signal also becomes less pronounced. The average longitudinal polarization of positrons decreases to $P_\parallel=8\%$, with the maximum polarization reaching $P^m_\parallel=40\%$ [Fig. \ref{Fig.phopos25fs} (d)]. In this scenario, to achieve a confidence level of $5\sigma$ for measuring vacuum birefringence, the required measurement time would need to be extended to 7 hours.  When the pulse duration is reduced, the effects of vacuum polarization remain detectable, but achieving a reasonable confidence level necessitates a relatively longer measurement time. 

\subsubsection{Laser intensity} We examine the fermionic signal with a laser intensity of $a_0=50$, which is comparable with current laser parameters at ELI Beamlines (1 PW pulses, repetition rate 10 Hz, pulse duration 30 fs). Assume the probe gamma photons are obtained by
linear Compton scattering of a linearly polarized laser pulse off an 15 GeV electron beam. The generated $\gamma$-photons within 0.02 mrad are highly polarized with $\xi_1=-0.87$ and have an average energy of $\omega_\gamma$=3GeV, with the energy spread $\Delta\omega_\gamma/\omega_\gamma=0.54$, see Fig. \ref{1pw_pho}. The yield of the gamma photons within 0.02 mrad  is $N_\gamma=0.44N_{e^-_0}$. Next, the probe photons propagate through a 1 PW laser pulse ($a_0=50$).  The polarization features of the created positrons are shown in Fig. \ref{1pw_pos}. The positrons are longitudinally polarized with average polarization of  $3.8\%$ and highest polarization up to $\sim14\%$ [Fig. \ref{1pw_pos}(b) and (f)]. The yield of positrons are $N_{e^+e^-}\approx 2.4 \times 10^5 \sim 0.25N_\gamma$ [Fig. \ref{1pw_pos} (a) and (e)]. Despite the decrease in total pair yield, the number of pairs within 10 mrad increases from $N_{e^+e^-}=1.3\times 10^6$ to $7.3\times 10^6$ due to the smaller deflection angle. The increase in positron density at small angles is beneficial for polarization measurement but is offset by the decrease in polarization. In order to detect VB at the 5$\sigma$ confidence level, the required positron number is $\tilde{N}_{e^+}= 1.4\times10^9$, corresponding to 200 shorts of measurement.  Considering the high repetition rate of the 1 PW laser is 10 Hz, the measurement time is 20 seconds, much smaller than 10 PW case.

The scaling law of measurement time and laser intensity is shown in Fig. \ref{scaling_law} (b).  With the increase of the laser intensity, the positron density increases monotonously, while the measurement time has an optimal at $a_0=150$ [Fig. \ref{scaling_law} (b)]. When the laser intensity increases from 100 to 150, the 
measurement time decreases from 3.5 to 2.9 hours due to the larger $\chi_\gamma$.
However, further increases in laser intensity lead to an increase in measurement time, as the probe photon undergoes pair production before attaining significant circular polarization.

\subsubsection{Collision angle of gamma and laser beams } 

The collision angle could also affect the pair yield and consequently induce a increase of measurement time. As shown in Fig. \ref{scaling_law} (c), the pairs yield decreases slightly from $N_{e^+e^-}=7.6\times 10^5$ to $N_{e^+e^-}=7.2\times 10^5$ with the increase of collision angle from $\theta_c=0^\circ$ to $\theta_c=20^\circ$.  When the positrons within $10$ mrad are collected for measurement, $t_{\text{Meas.}}$ increases significantly from 2.9 to 257 hours. However, if the detection angle of positrons is rotated with collision angle, the measurement time remains $\sim 3$hours, which is robust against the fluctuation of $\theta_c$ .

}

\section{Conclusion}

Concluding, we analyzed a setup for a high-energy VP measurement using a 10 PW laser system with 1 GeV linearly polarized $\gamma$ probe photons, with a newly developed complete QED Monte Carlo simulation method for describing vacuum polarization in the high-energy limit. Deviating from the conventional photonic signal of VP, we identified the fermionic signal of VB in the positron polarization that is free from disturbances caused by secondary emissions, and more feasible for VB detection. In our scheme, previously avoided real pairs are employed as a better source for detecting VB, providing a novel method for probing quantum vacuum nonlinearity. {\color{black} The fermionic signal remains robust against experimental fluctuations, enabling a $5\sigma$ confidence level within a few hours.

In addition, the high polarization and density of gamma photons allows for a single-shot measurement for vacuum polarization, achieving an $8\sigma$ confidence level. The revealed polarization feature of positrons provides an alternative way of measuring vacuum birefringence.  As a by-product, our scheme supplies 
a well-collimated ($\sim$0.05 mrad), dense ($\sim 2.7\times 10^5$) and highly circularly polarized gamma-ray beam with an average polarization reaching up to $60\%$, as well as a dense ($7.5\times 10^5$) longitudinally polarized positrons with a highest polarization of $\sim 70\%$ via QED loop effects. 
Besides of the potential application in detecting vacuum birefringence, such polarized particles are highly demanded in studies of fundamental physics  and related applications, in particular, in  nuclear physics, astrophysics, and high-precision high-energy physics at accelerators, including parity violation,  photon-photon scattering, and photoproduction of mesons. }

\section*{Acknowledgement}
We gratefully acknowledge helpful discussions with Prof. Y.-F.~Li. This work is supported by the National Natural Science Foundation of China (Grants No. 12074262) and the National Key R\&D Program of China (Grant No. 2021YFA1601700).

\appendix 

{\color{black}
\section{The QED treatment of vacuum polarization}\label{A1}

According to the QED loop calculation in \cite{torgrimsson2021loops}, the $O\left(\alpha^{0}\right)$-order loop contribution is
 \begin{align*}
P^{0} & =\frac{1}{2}\left(1+\boldsymbol{\xi}_{i}\cdot\boldsymbol{\xi}'\right),\numberthis
\end{align*}
where the initial and final photon polarizations are represented by the Stokes parameters $\boldsymbol{\xi}_i$ and $\boldsymbol{\xi}_f$, respectively. The $O\left(\alpha\right)-$order loop contribution via the interference diagram in Fig. \ref{Fig.diagram}(b) reads
\[
P^{1}=\left\langle P^{L}\right\rangle +P_{0}^{L}\cdot n_{0}+P_{1}^{L}\cdot n_{1}+n_{1}\cdot P_{10}^{L}\cdot n_{0}.\numberthis
\]
The sum of these contributions is 
\begin{align} 
\label{loop}
P^{L}&=P^{L}_{VD}+P^{L}_{VB}\\
P^{L}_{VD}&=\frac{1}{2}\left(1-\int_{0}^{\omega}d\varepsilon{C}_{p}\left[\int_{z_{p}}^{\infty}dx\textrm{K}_{\frac{1}{3}}\left(x\right)\right.\right.\\\nonumber
&\left.\left.+\frac{\varepsilon^{2}+\varepsilon_{+}^{2}}{\varepsilon_{+}\varepsilon}\textrm{K}_{\frac{2}{3}}\left(u'\right)-\textrm{K}_{\frac{2}{3}}\left(z_{p}\right)\mathbf{\hat{e}_{3}}\cdot\boldsymbol{\xi}_{i}\right]\right)\\\nonumber
&+\boldsymbol{\xi}'\left[\left(1-\int_{0}^{\omega}d\varepsilon{C}_{p}\left[\int_{z_{p}}^{\infty}dx\textrm{K}_{\frac{1}{3}}\left(x\right)+\frac{\varepsilon^{2}+\varepsilon_{+}^{2}}{\varepsilon_{+}\varepsilon}\textrm{K}_{\frac{2}{3}}\left(z_{p}\right)\right]\right)\boldsymbol{\xi}_{i}\right.\\\nonumber
&\left.+\int_{0}^{\omega}d\varepsilon{C}_{p}\textrm{K}_{\frac{2}{3}}\left(z_{p}\right)\mathbf{\hat{e}_{3}}\right],\\
P^{L}_{VB}&=\frac{\alpha m^{2}}{2\omega^{2}}\Delta t\int d\varepsilon\frac{\textrm{Gi}'\left(\rho\right)}{\rho}\boldsymbol{\xi}'\cdot\varepsilon_{3ij}\cdot\boldsymbol{\xi}_{i}\nonumber\\
 & =\frac{\alpha m^{2}}{2\omega^{2}}\Delta t\int d\varepsilon\frac{\textrm{Gi}'\left(\rho\right)}{\rho}\left(\xi_{1}^{'}\xi_{i2}-\xi_{2}^{'}\xi_{i1}\right),\numberthis
\end{align}
where $C_{p}=\frac{\alpha m^{2}}{\sqrt{3}\pi\omega^{2}}$,  $z_p=\frac{2}{3\chi_\gamma}\frac{\omega^2}{\varepsilon_+\varepsilon}$, {$\chi_\gamma=|F_{\mu\nu}k^\nu|/mF_{cr}$} the strong-field quantum parameter, $\varepsilon_+$ and $\varepsilon$ are the energy of produced positrons and electrons, respectively,  
 $\rho=\frac{1}{\left[\delta\left(1-\delta\right)\chi_{\gamma}\right]^{2/3}}$, $\delta=\varepsilon/\omega$,  and Gi$'(x)$ is the Scorer prime function.

The first term of Eq. (\ref{loop}) $P^{L}_{VD}$, stemming from the imaginary part of of the polarization operator describes VD. While the second term $P^{L}_{VB}$ is associated with the real part of the polarization operator, and induces VB.  

When the pair production is negligible, the loop probability of all orders can be resummed into a time-ordered exponential \cite{torgrimsson2021loops}:
\begin{align*}\label{nbw}
P&=\frac{1}{2}\boldsymbol{\xi}'\cdot e^{-4\left\langle P^{BW}\right\rangle }\left[\left(\epsilon_{0}\epsilon_{0}+\epsilon_{3}\epsilon_{3}\right)\cosh\nu+\left(\epsilon_{0}\epsilon_{3}+\epsilon_{3}\epsilon_{0}\right)\sinh\nu\right.\\
&\left.+\left(\epsilon_{1}\epsilon_{1}+\epsilon_{2}\epsilon_{2}\right)\cos\varphi+\left(\epsilon_{1}\epsilon_{2}-\epsilon_{2}\epsilon_{1}\right)\text{\ensuremath{\sin}}\varphi\right]\boldsymbol{\xi}_{i}\\
&=\frac{e^{-W_{P}}}{2}\left[\left(\cosh v+\xi_{i3}\sinh v\right)+\xi_{1}^{'}\left(\xi_{i1}\cos\varphi+\xi_{i2}\sin\varphi\right)\right.\\
&+\left.\xi_{2}^{'}\left(-\xi_{i1}\sin\varphi+\xi_{i2}\cos\varphi\right)+\xi_{3}^{'}\left(\sinh v+\xi_{i3}\cosh v\right)\right],\numberthis
\end{align*}
where $W_{P}=4\left\langle P^{BW}\right\rangle$ is the total pair production probability, and 
\begin{align}\label{varphi}
\varphi & =\frac{\alpha m^{2}}{\omega^{2}}\Delta t\int d\varepsilon\frac{\textrm{Gi}'\left(\rho\right)}{\rho}, \quad v=-\frac{\alpha m^{2}}{\omega^{2}}\Delta t\int d\varepsilon\frac{\textrm{Ai}'\left(\rho\right)}{\rho},
\end{align}
with Ai$'(x)$ being the Airy prime function.
The final stokes parameters for the remaining probe photons read
\begin{align*}\label{xif0}
\xi_{1}^{f}&=\frac{\xi{}_{1}\cos\varphi+\xi{}_{2}\sin\varphi}{\cosh v+\xi_{3}\sinh v},\\
\xi_{2}^{f}&=\frac{-\xi{}_{1}\sin\varphi+\xi{}_{2}\cos\varphi}{\cosh v+\xi_{3}\sinh v},\\
\xi_{3}^{f}&=\frac{\sinh v+\xi_{3}\cosh v}{\cosh v+\xi_{3}\sinh v}.\numberthis
\end{align*}
The photon number at a distance $l$ takes the form 
\begin{align}\label{n}
N\left(l\right)&=e^{-W_{P}}\left(\cosh v+\xi_{3}\sinh v\right)N\left(0\right).
\end{align}
 The equations (\ref{xif0}) and (\ref{n})  coincide with Eq.~(15.20) given in \cite{baier1998electromagnetic}.
The average polarization of a photon ensemble defined as $\bar{\xi}^{f}=\xi_{L}^{f}W^{L}$, with the final polarization state of photons $\xi_{L}^{f}=(\xi_1^f,\xi_2^f,\xi_3^f)$, and the loop probability $W^{L}=e^{-W_{P}}\left(\cosh v+\xi_{3}\sinh v\right)$, reads 
\begin{align*}\label{average}
\left(\begin{array}{c}
\bar{\xi}_{0}\\
\bar{\xi}{}_{1}\\
\bar{\xi}{}_{2}\\
\bar{\xi}{}_{3}
\end{array}\right)&=e^{-W_{P}}\left(\begin{array}{cccc}
\cosh\nu & 0 & 0 & \sinh\nu\\
0 & \cos\varphi & \sin\varphi & 0\\
0 & -\sin\varphi & \cos\varphi & 0\\
\sinh\nu & 0 & 0 & \cosh\nu
\end{array}\right)\left(\begin{array}{c}
\xi_{0}\\
\xi_{1}\\
\xi_{2}\\
\xi_{3}
\end{array}\right),\numberthis
\end{align*}
which coincides with Eq. (11) of Ref. \cite{bragin2017high}. However, rather than averaging over the survived ones as in the present work, the polarization defined in Ref. \cite{bragin2017high} is obtained by averaging over the initial photon number:
\begin{equation}\label{average_stokes}
\ensuremath{\bar{\xi}^{f}}=\xi_{L}^{f}W^{L}=\frac{N_{\uparrow}^{NP}-N_{\downarrow}^{NP}}{N_{\uparrow}^{NP}+N_{\downarrow}^{NP}}\cdot\frac{N^{NP}}{N^{NP}+N^{P}}=\frac{N_{\uparrow}^{NP}-N_{\downarrow}^{NP}}{N^{NP}+N^{P}},
\end{equation}
where $N^{NP}=N_{\uparrow}^{NP}+N_{\downarrow}^{NP}$ is the number of photons that are survived from pair production, with $N_{\uparrow}^{NP}$ and $N_{\downarrow}^{NP}$ being the number of photons with final polarization $\boldsymbol{\xi}'=\pm \boldsymbol{\xi}^f$ and $\boldsymbol{\xi}^f=(\xi^f_1,\xi^f_2,\xi^f_3)/\sqrt{\xi^{f2}_1+\xi^{f2}_2+\xi^{f2}_3}$, and $N^{P}$ is the number of photons that decay into pairs.
Therefore, the average polarization defined by Eq. (\ref{average_stokes}) is by the factor $\frac{N^{NP}}{N^{NP}+N^{P}}$ smaller than the polarization of the survived photons in the final state. For small $\chi_\gamma\ll 1$,  the difference between the definitions of the photon final polarization is negligible.

\section{Monte-Carlo simulation method for vacuum birefringence and dichroism}\label{A2}

In this section, we present the spin- and polarization-resolved Monte-Carlo method for the tree-process (nonlinear Breit-Wheeler) and the loop-process (vacuum polarization). In our Monte Carlo code, at each simulation step $\Delta t$,  the pair production is determined by the total pair production probability, and the positron energy and polarization by the spin-resolved spectral probability  \cite{chen2022electron}, using the common algorithms \cite{zhuang2023laser,dai2022photon,chen2019polarized,li2020production,gonoskov2015extended,
ridgers2014modelling,elkina2011qed,green2015simla}. If the pair production event is rejected, the photon polarization state is determined by the photon-polarization dependent loop probability $w^{NP}$. 

\subsection{Spin- and polarization-resolved pair production probability}

The pair production probability including all the polarization and spin characteristics takes the form \cite{chen2022electron,zhuang2023laser}
\begin{align}\nonumber\label{PPP2}
dW^{P}\left(\bm{\xi},\bm{\zeta}_{-},\bm{\zeta}_{+}\right)&=\frac{1}{2}\left(dW_{11}+dW_{22}\right)+\frac{\xi_{1}}{2}\left(dW_{11}-dW_{22}\right)\\\nonumber
&-i\frac{\xi_{2}}{2}\left(dW_{21}-dW_{12}\right)+\frac{\xi_{3}}{2}\left(dW_{11}-dW_{22}\right)\\
&=\frac{1}{2}\left(G_{0}+\xi_{1}G_{1}+\xi_{2}G_{2}+\xi_{3}G_{3}\right),
\end{align}
where
\begin{align*}\nonumber
G_{0}&=\frac{{C}_{p}}{2}d\varepsilon\Bigg\{\left\{ \int_{z_{p}}^{\infty}dx\textrm{K}_{\frac{1}{3}}\left(x\right)+\frac{\varepsilon_{+}^{2}+\varepsilon^{2}}{\varepsilon_{+}\varepsilon}\textrm{K}_{\frac{2}{3}}\left(z_{p}\right)\right\} \\&+\left\{ \int_{z_{p}}^{\infty}dx\textrm{K}_{\frac{1}{3}}\left(x\right)-2\textrm{K}_{\frac{2}{3}}\left(z_{p}\right)\right\} \left(\bm{\zeta}_{-}\cdot\bm{\zeta}_{+}\right)\\&+\left[\frac{\omega}{\varepsilon_{+}}\left(\bm{\zeta}_{+}\cdot\mathbf{b}\right)-\frac{\omega}{\varepsilon}\left(\bm{\zeta}_{-}\cdot\mathbf{b}\right)\right]\textrm{K}_{\frac{1}{3}}\left(z_{p}\right)\\&+\left\{ \frac{\varepsilon_{+}^{2}+\varepsilon^{2}}{\varepsilon\varepsilon_{+}}\int_{z_{p}}^{\infty}dx\textrm{K}_{\frac{1}{3}}\left(x\right)\right.\\&-\left.\left.\frac{\left(\varepsilon_{+}-\varepsilon\right)^{2}}{\varepsilon\varepsilon_{+}}\textrm{K}_{\frac{2}{3}}\left(z_{p}\right)\right\} \left(\bm{\zeta}_{-}\cdot\mathbf{\hat{v}}\right)\left(\bm{\zeta}_{+}\cdot\mathbf{\hat{v}}\right)\right\},
\end{align*}
\begin{align*}\nonumber
G_{1}&=\frac{{C}_{p}}{2}d\varepsilon\Bigg\{-\frac{\varepsilon_{+}^{2}-\varepsilon^{2}}{2\varepsilon_{+}\varepsilon}\textrm{K}_{\frac{2}{3}}\left(z_{p}\right)\mathbf{\hat{v}}\cdot\left[\bm{\zeta}_{+}\times\bm{\zeta}_{-}\right]\\&+\left[\frac{\omega}{\varepsilon}\left(\bm{\zeta}_{+}\cdot\mathbf{s}\right)-\frac{\omega}{\varepsilon_{+}}\left(\bm{\zeta}_{-}\cdot\mathbf{s}\right)\right]\textrm{K}_{\frac{1}{3}}\left(z_{p}\right)\\&-\frac{\omega^{2}}{2\varepsilon_{+}\varepsilon}\int_{z_{p}}^{\infty}dx\textrm{K}_{\frac{1}{3}}\left(x\right)\left\{ \left(\bm{\zeta}_{-}\cdot\mathbf{b}\right)\left(\bm{\zeta}_{+}\cdot\mathbf{s}\right)+\left(\bm{\zeta}_{-}\cdot\mathbf{s}\right)\left(\bm{\zeta}_{+}\cdot\mathbf{b}\right)\right\} \Bigg\},
\end{align*}
\begin{align*}\nonumber
G_{2}&=\frac{{C}_{p}}{2}d\varepsilon\Bigg\{-\frac{\omega^{2}}{2\varepsilon_{+}\varepsilon}\textrm{K}_{\frac{1}{3}}\left(z_{p}\right)\left[\mathbf{s}\cdot(\bm{\zeta}_{-}\times\bm{\zeta}_{+})\right]\\&+\left(\frac{\omega}{\varepsilon_{+}}\int_{z_{p}}^{\infty}dx\textrm{K}_{\frac{1}{3}}\left(x\right)+\frac{\varepsilon_{+}^{2}-\varepsilon^{2}}{\varepsilon_{+}\varepsilon}\textrm{K}_{\frac{2}{3}}\left(z_{p}\right)\right)\left(\bm{\zeta}_{+}\cdot\mathbf{\hat{v}}\right)\\&+\left(\frac{\omega}{\varepsilon}\int_{z_{p}}^{\infty}dx\textrm{K}_{\frac{1}{3}}\left(x\right)-\frac{\varepsilon_{+}^{2}-\varepsilon^{2}}{\varepsilon_{+}\varepsilon}\textrm{K}_{\frac{2}{3}}\left(z_{p}\right)\right)\left(\bm{\zeta}_{-}\cdot\mathbf{\hat{v}}\right)\\&-\frac{\varepsilon_{+}^{2}-\varepsilon^{2}}{2\varepsilon_{+}\varepsilon}\textrm{K}_{\frac{1}{3}}\left(z_{p}\right)\left[\left(\bm{\zeta}_{-}\cdot\mathbf{\hat{v}}\right)\left(\bm{\zeta}_{+}\cdot\mathbf{b}\right)+\left(\bm{\zeta}_{-}\cdot\mathbf{b}\right)\left(\bm{\zeta}_{+}\cdot\mathbf{\hat{v}}\right)\right]\Bigg\},
\end{align*}
\begin{align}\nonumber \label{PRB_PHO}
G_{3}&=\frac{{C}_{p}}{2}d\varepsilon\Bigg\{-\textrm{K}_{\frac{2}{3}}\left(z_{p}\right)+\frac{\varepsilon_{+}^{2}+\varepsilon^{2}}{2\varepsilon_{+}\varepsilon}\textrm{K}_{\frac{2}{3}}\left(z_{p}\right)\left(\bm{\zeta}_{-}\cdot\bm{\zeta}_{+}\right)\\\nonumber
&+\left[-\frac{\omega}{\varepsilon}\left(\bm{\zeta}_{+}\cdot\mathbf{b}\right)+\frac{\omega}{\varepsilon_{+}}\left(\bm{\zeta}_{-}\cdot\mathbf{b}\right)\right]\textrm{K}_{\frac{1}{3}}\left(z_{p}\right)\\\nonumber
&-\frac{\left(\varepsilon_{+}-\varepsilon\right)^{2}}{2\varepsilon_{+}\varepsilon}\textrm{K}_{\frac{2}{3}}\left(z_{p}\right)\left(\bm{\zeta}_{-}\cdot\mathbf{\hat{v}}\right)\left(\bm{\zeta}_{+}\cdot\mathbf{\hat{v}}\right)\\
&+\frac{\omega^{2}}{2\varepsilon_{+}\varepsilon}\int_{z_{p}}^{\infty}dx\textrm{K}_{\frac{1}{3}}\left(x\right)\left[\left(\bm{\zeta}_{-}\cdot\mathbf{b}\right)\left(\bm{\zeta}_{+}\cdot\mathbf{b}\right)-\left(\bm{\zeta}_{-}\cdot\mathbf{s}\right)\left(\bm{\zeta}_{+}\cdot\mathbf{s}\right)\right]\Bigg\}.
\end{align}\\
Here $\mathbf{\hat{v}}$ is the unit vector along velocity of the produced electron, $\mathbf{s}$ the unit vector along the transverse component of electron acceleration, and $\mathbf{b}=\mathbf{\hat{v}}\times \mathbf{s}$. The 3-vector $\bm{\xi}=\left(\xi_{1},\xi_{2},\xi_{3}\right)$ is the Stokes parameter of the incoming photon, $\omega$  the photon energy and $\varepsilon_{+}$ and $\varepsilon_{-}$ are the energy of the created positron and electron, respectively.\\

\subsubsection*{Spin quantization axis for the produced electron}

 After taking the sum over positron polarizations \cite{chen2022electron}:
\begin{align*}\nonumber
 d\widetilde{W}^{p} \left(\bm{\xi},\bm{\zeta}_-\right) & =\frac{1}{2}\left(\widetilde{G}_{0}+\xi_{1}\widetilde{G}_{1}+\xi_{2}\widetilde{G}_{2}+\xi_{3}\widetilde{G}_{3}\right),
\end{align*}
\begin{align*}\nonumber
\widetilde{G}_{0}&={C}_{p}d\varepsilon\Bigg\{\int_{z_{p}}^{\infty}dx\textrm{K}_{\frac{1}{3}}\left(x\right)+\frac{\varepsilon_{+}^{2}+\varepsilon^{2}}{\varepsilon_{+}\varepsilon}\textrm{K}_{\frac{2}{3}}\left(z_{p}\right)\\&-\frac{\omega}{\varepsilon}\left(\bm{\zeta}_{-}\cdot\mathbf{b}\right)\textrm{K}_{\frac{1}{3}}\left(z_{p}\right)\Bigg\}
\end{align*}
\begin{align*}\nonumber
\widetilde{G}_3 & ={C}_pd\varepsilon\Bigg\{-\textrm{K}_{\frac{2}{3}}\left(z_{p}\right)
 +\frac{\omega}{\varepsilon_{+}}\left(\bm{\zeta}_{-}\cdot\mathbf{b}\right)\textrm{K}_{\frac{1}{3}}\left(z_{p}\right)\Bigg\}
\end{align*}
\begin{align*}\nonumber
\widetilde{G}_1 & =-{C}_pd\varepsilon\frac{\omega}{\varepsilon_{+}}\left(\bm{\zeta}_{-}\cdot\mathbf{s}\right)\textrm{K}_{\frac{1}{3}}\left(z_{p}\right)
\end{align*}
\begin{align}\label{PPP3}
\widetilde{G}_2 & ={C}_pd\varepsilon\Bigg\{ \left(\frac{\omega}{\varepsilon}\int_{z_{p}}^{\infty}dx\textrm{K}_{\frac{1}{3}}\left(x\right)-\frac{\varepsilon_{+}^{2}-\varepsilon^{2}}{\varepsilon_{+}\varepsilon}\textrm{K}_{\frac{2}{3}}\left(z_{p}\right)\right)\left(\bm{\zeta}_{-}\cdot\mathbf{\hat{v}} \right)\Bigg\},
\end{align}
which can be rewritten in the form \cite{chen2022electron}
\begin{align}\nonumber
d\widetilde{W}^{p}\left(\bm{\xi},\bm{\zeta}_{-}\right)&=\frac{1}{2}\left(a_{-}+\bm{\zeta}_{-}\cdot\bm{b}_{-}\right)\\\nonumber
a_{-}&={C}_{p}d\varepsilon\left[\int_{z_{p}}^{\infty}dx\textrm{K}_{\frac{1}{3}}\left(x\right)\right.\\\nonumber
&+\left.\frac{\varepsilon_{+}^{2}+\varepsilon^{2}}{\varepsilon_{+}\varepsilon}\textrm{K}_{\frac{2}{3}}\left(z_{p}\right)-\xi_{3}\textrm{K}_{\frac{2}{3}}\left(z_{p}\right)\right]\\\nonumber
\bm{b}_{-}&=-{C}_{p}d\varepsilon\left\{ \xi_{1}\frac{\omega}{\varepsilon_{+}}\mathbf{s}\textrm{K}_{\frac{1}{3}}\left(z_{p}\right)+\left(\frac{\omega}{\varepsilon}-\xi_{3}\frac{\omega}{\varepsilon_{+}}\right)\mathbf{b}\textrm{K}_{\frac{1}{3}}\left(z_{p}\right)\right.\\\nonumber
&+\left.\left[-\frac{\omega}{\varepsilon}\int_{z_{p}}^{\infty}dx\textrm{K}_{\frac{1}{3}}\left(x\right)+\frac{\varepsilon_{+}^{2}-\varepsilon^{2}}{\varepsilon_{+}\varepsilon}\textrm{K}_{\frac{2}{3}}\left(z_{p}\right)\right]\xi_{2}\mathbf{\hat{v}}\right\}.\\
\end{align}
The final polarization vector of the produced electron resulting from the scattering process itself is $\bm{\zeta}^-_f=\frac{\bm{b}_-}{a_-}$, which determines the spin quantization axis for the produced electron  $\bm{\zeta}_{f}^{-}$: $\bm{n}^-=\bm{\zeta}_f^-/|\bm{\zeta}_f^-|$.\\

\subsubsection*{Spin quantization axis for the produced positron}

 After taking the sum over electron polarizations we obtain \cite{chen2022electron}:
\begin{align*}\nonumber
 d\overline{W}^{p} \left(\bm{\xi},\bm{\zeta}_+\right)& =\frac{1}{2}\left(\overline{G}_{0}+\xi_{1}\overline{G}_{1}+\xi_{2}\overline{G}_{2}+\xi_{3}\overline{G}_{3}\right),
\end{align*}
\begin{align*}\nonumber
\overline{G}_{0}&={C}_{p}d\varepsilon\Bigg\{\int_{z_{p}}^{\infty}dx\textrm{K}_{\frac{1}{3}}\left(x\right)+\frac{\varepsilon_{+}^{2}+\varepsilon^{2}}{\varepsilon_{+}\varepsilon}\textrm{K}_{\frac{2}{3}}\left(z_{p}\right)\\&+\frac{\omega}{\varepsilon}_{+}\left(\bm{\zeta}_{+}\cdot\mathbf{b}\right)\textrm{K}_{\frac{1}{3}}\left(z_{p}\right)\Bigg\}
\end{align*}
\begin{align*}\nonumber
\overline{G}_3 & ={C}_pd\varepsilon\Bigg\{-\textrm{K}_{\frac{2}{3}}\left(z_{p}\right)
 -\frac{\omega}{\varepsilon}\left(\bm{\zeta}_{+}\cdot\mathbf{b}\right)\textrm{K}_{\frac{1}{3}}\left(z_{p}\right)\Bigg\}
\end{align*}
\begin{align*}\nonumber
\overline{G}_1 & ={C}_pd\varepsilon\frac{\omega}{\varepsilon}\left(\bm{\zeta}_{+}\cdot\mathbf{s}\right)\textrm{K}_{\frac{1}{3}}\left(z_{p}\right)
\end{align*}
\begin{align}\label{PPP3}
\overline{G}_2 & ={C}_pd\varepsilon\Bigg\{ \left(\frac{\omega}{\varepsilon}_{+}\int_{z_{p}}^{\infty}dx\textrm{K}_{\frac{1}{3}}\left(x\right)+\frac{\varepsilon_{+}^{2}-\varepsilon^{2}}{\varepsilon_{+}\varepsilon}\textrm{K}_{\frac{2}{3}}\left(z_{p}\right)\right)\left(\bm{\zeta}_{+}\cdot\mathbf{\hat{v}} \right)\Bigg\},
\end{align}
which can also be written as \cite{chen2022electron}
\begin{align}\nonumber
d\overline{W}^{p}\left(\bm{\xi},\bm{\zeta}_{+}\right)&=\frac{1}{2}\left(a_{+}+\bm{\zeta}_{+}\cdot\bm{b}_{+}\right)\\\nonumber
a_{+}&={C}_{p}d\varepsilon\left[\int_{z_{p}}^{\infty}dx\textrm{K}_{\frac{1}{3}}\left(x\right)\right.\\&+\left.\frac{\varepsilon_{+}^{2}+\varepsilon^{2}}{\varepsilon_{+}\varepsilon}\textrm{K}_{\frac{2}{3}}\left(z_{p}\right)-\xi_{3}\textrm{K}_{\frac{2}{3}}\left(z_{p}\right)\right]\\\nonumber
\bm{b}_{+}&={C}_{p}d\varepsilon\left\{ \xi_{1}\textrm{K}_{\frac{1}{3}}\left(z_{p}\right)\frac{\omega}{\varepsilon}\mathbf{s}\right.\\\nonumber
&+\left(\frac{\omega}{\varepsilon}_{+}-\xi_{3}\overline{C}_{0}d\varepsilon\frac{\omega}{\varepsilon}\right)\mathbf{b}\textrm{K}_{\frac{1}{3}}\left(z_{p}\right)\\\nonumber
&+\left.\xi_{2}\mathbf{\hat{v}}\left(\frac{\omega}{\varepsilon}_{+}\int_{z_{p}}^{\infty}dx\textrm{K}_{\frac{1}{3}}\left(x\right)+\frac{\varepsilon_{+}^{2}-\varepsilon^{2}}{\varepsilon_{+}\varepsilon}\textrm{K}_{\frac{2}{3}}\left(z_{p}\right)\right)\right\}.\\
\end{align}

The final polarization vector of the produced positron resulting from the scattering process itself is $\bm{\zeta}^+_f=\frac{\bm{b}_+}{a_+}$, which determines the spin quantization axis for the produced positron: $\bm{n}^+=\bm{\zeta}^+_f/|\bm{\zeta}^+_f|$.

After taking the sum over positron and electron polarizations, we get the spin unresolved pair production probability:
\begin{align}\label{WP0}
dW_{T}^{P}&\left(\bm{\xi}\right)=a_+.
\end{align}
\subsection{Polarization-resolved no-pair production probability}
If a pair production event is rejected, the photon polarization  should also change due to the dependency of the no-pair production probability on the photon polarization:
\begin{align}\nonumber \label{PRB_NP}
W^{NP}\left(\boldsymbol{\xi},\boldsymbol{\xi}'\right)&=\frac{1}{2}\left(c^{NP}+\boldsymbol{d}^{NP}\cdot\boldsymbol{\xi}'\right)\\\nonumber c^{NP}&=1-\int_{0}^{\omega}a_{+}d\varepsilon\Delta t\\\nonumber\bm{d}^{NP}&=\boldsymbol{\xi}\left(1-\int_{0}^{\omega}d\varepsilon{C}_{p}\left[\int_{z_{p}}^{\infty}dx\textrm{K}_{\frac{1}{3}}\left(x\right)\right.\right.\\&+\left.\left.\frac{\varepsilon_{+}^{2}+\varepsilon^{2}}{\varepsilon\varepsilon_{+}}\textrm{K}_{\frac{2}{3}}\left(z_{p}\right)\right]\Delta t\right)+\int_{0}^{\omega}d\varepsilon{C}_{p}\mathbf{\hat{e}_{3}}\textrm{K}_{\frac{2}{3}}\left(z_{p}\right)\Delta t.
\end{align}
where $\mathbf{\hat{e}_3}=(0,0,1)$. The final polarization state of the photon after the no-pair-production step becomes $\bm{\xi}_f^{NP}=\bm{d}^{NP}/c^{NP}$, which defines a quantization axis for photon polarization: $\bm{n}^{NP}=\bm{\xi}_f^{NP}/|\bm{\xi}_f^{NP}|$.

\subsection{Algorithm of event generation}

1. \textit{Update photon polarization } At each time step, the photon polarization needs to be updated with local acceleration.

(1) Calculate the instantaneous polarization basis  vectors $\mathbf{e}_{1}=\mathbf{s}-\left(\mathbf{n\cdot s}\right)\mathbf{s}$ and $\mathbf{e}_{2}=\mathbf{n\times s}$, with unit vectors of electron acceleration $\mathbf{s}$ and photon propagation direction $\mathbf{n}$. 

(2) Update the photon Stokes parameters
\begin{align*}
\xi_{1}'&=\xi_{1}\cos\left(2\psi\right)-\xi_{3}\sin\left(2\psi\right),\\
\xi_{2}'&=\xi_{2},\\
\xi_{3}'&=\xi_{1}\sin\left(2\psi\right)+\xi_{3}\cos\left(2\psi\right),
\end{align*}
where $\psi$ is the angle between the new and old basis vectors.
 
2. \textit{Decide pair production event}: At each simulation step, the pair production and the electron (positron) energy are determined by the probability of Eq. (\ref{WP0}) with the updated stokes parameters, using the common stochastic procedure.

(1) Generate two random numbers $r_1,r_2\in [0,1]$ with uniform probability.

(2) Compute the pair production probability $P(r_1)=dW_T^P(\bm{\xi}, r_1\omega)\Delta t$  for the given initial photon polarization $\bm{\xi}$,  electron energy  $\varepsilon=r_1\omega$ and positron energy $\varepsilon_+=(1-r_1)\omega$.

(3) If $r_2<P(r_1)$, an $e^+e^-$ pair is created. Otherwise, reject.\\

3. \textit{Decide the polarization of outgoing particles}: \\

\textbf{Case 1}: $P(r_1)>r_2$: pair production occurs. After each pair production, the spin of the produced electron (positron) is either parallel or antiparallel to $\bm{n}^-$ ($\bm{n}^+$)  using the stochastic procedure with another random number $r_3\in[0,1]$. With the given $\varepsilon_-$, $\varepsilon_+$ and photon polarization $\bm{\xi}$, compute the pair production probability $P_{\bm{\zeta_-}\bm{\zeta_+}}=dW^P(\bm{\xi},\bm{\zeta_-},\bm{\zeta_+})\Delta t$ with $\{\bm{\zeta_-},\bm{\zeta_+}\}\in \{\uparrow,\downarrow\}$ indicating parallel or antiparallel with respective quantization axis.

(1)  If $r_3<P_{\downarrow\downarrow}$, the electron is spin down with respect to $\bm{n}^-$ and the positron is  spin down with respect to $\bm{n}^+$, i.e. $\bm{\zeta}_-=-\bm{n}^-$, $\bm{\zeta}_+=-\bm{n}^+$

(2) If $P_{\downarrow\downarrow}<r_3<P_{\downarrow\downarrow}+P_{\downarrow\uparrow}$, $\bm{\zeta_-}=-\bm{n}^-$ and $\bm{\zeta_+}=\bm{n}^+$.

(3) If $P_{\downarrow\downarrow}+P_{\downarrow\uparrow}<r_3<P_{\downarrow\downarrow}+P_{\downarrow\uparrow}+P_{\uparrow\downarrow}$, $\bm{\zeta_-}=\bm{n}^-$ and $\bm{\zeta_+}=-\bm{n}^+$.

(4) If $P_{\downarrow\downarrow}+P_{\downarrow\uparrow}+P_{\uparrow\downarrow}<r_3<P_{\downarrow\downarrow}+P_{\downarrow\uparrow}+P_{\uparrow\downarrow}+P_{\uparrow\uparrow}$, $\bm{\zeta_-}=\bm{n}^-$ and $\bm{\zeta_+}=\bm{n}^+$.

\textbf{Case 2}: $P(r_1)<r_2$: pair production is rejected. The photon polarization state collapses into one of its basis states defined with respect to $\bm{n}^{NP}$.

(1) Generate another random number $r_4\in[0,1]$.

(2) Compute the no-pair-production probability $P_{\bm{\xi}'}=W^{NP}\left(\bm{\xi},\bm{\xi}'\right)$ for a given initial photon polarization $\bm{\xi}$. Here $\bm{\xi}'\in\{\uparrow,\downarrow\}$ indicates spin parallel or antiparallel with $\bm{n}^{NP}$.

(3) If $P_\uparrow/\left(P_\uparrow+P_\downarrow\right)>r_4$, $\bm{\xi}'=\bm{n}^{NP}$. Otherwise, $\bm{\xi}'=-\bm{n}^{NP}$.

In the above algorithm, the pair spin (photon polarization) is determined by the spin-resolved (photon-polarization-resolved) probabilities according to the stochastic algorithm and instantaneously collapses into one of its basis states defined with respect to the instantaneous quantization axis (SQA). Alternatively, one could set the pairs in a mixed spin state $\bm{\zeta}_\pm'=\bm{\zeta}^\pm_f$ or photon polarization $\bm{\xi}=\bm{\xi}_f^{NP}$ in the case of no-pair production.

4. \textit{Rotate the photon polarization}

Calculate the instantaneous retarded phase induced by vacuum birefringence use Eq. (\ref{varphi}), and update $\xi_1$ and $\xi_2$ following \cite{torgrimsson2021loops,bragin2017high,dinu2014vacuum}:
\begin{align}\label{VB_rotate}
\left(\begin{array}{c}
\xi{}_{1}^{f}\\
\xi{}_{2}^{f}
\end{array}\right)	=\left(\begin{array}{cc}
\cos\varphi & \sin\varphi\\
-\sin\varphi & \cos\varphi
\end{array}\right)\left(\begin{array}{c}
\xi_{1}\\
\xi_{2}
\end{array}\right).
\end{align}

\subsection{Benchmark of our simulation method}

We have demonstrated the no-pair production probability used in our code corresponds to the loop probability, with which Eq. (15.20) in Ref. \cite{baier1998electromagnetic} and Eq. (11) in Ref. \cite{bragin2017high} can be reproduced. To further benchmark the accuracy of our code, we have plotted the phase variation induced by vacuum birefringence and final stokes parameters for different parameters.  With the parameters used in Ref. \cite{bragin2017high}, our results are in good agreement with Fig.4 and Fig. 5 in Ref. \cite{bragin2017high}. 

\begin{figure}[htp]
	\includegraphics[width=0.48\textwidth]{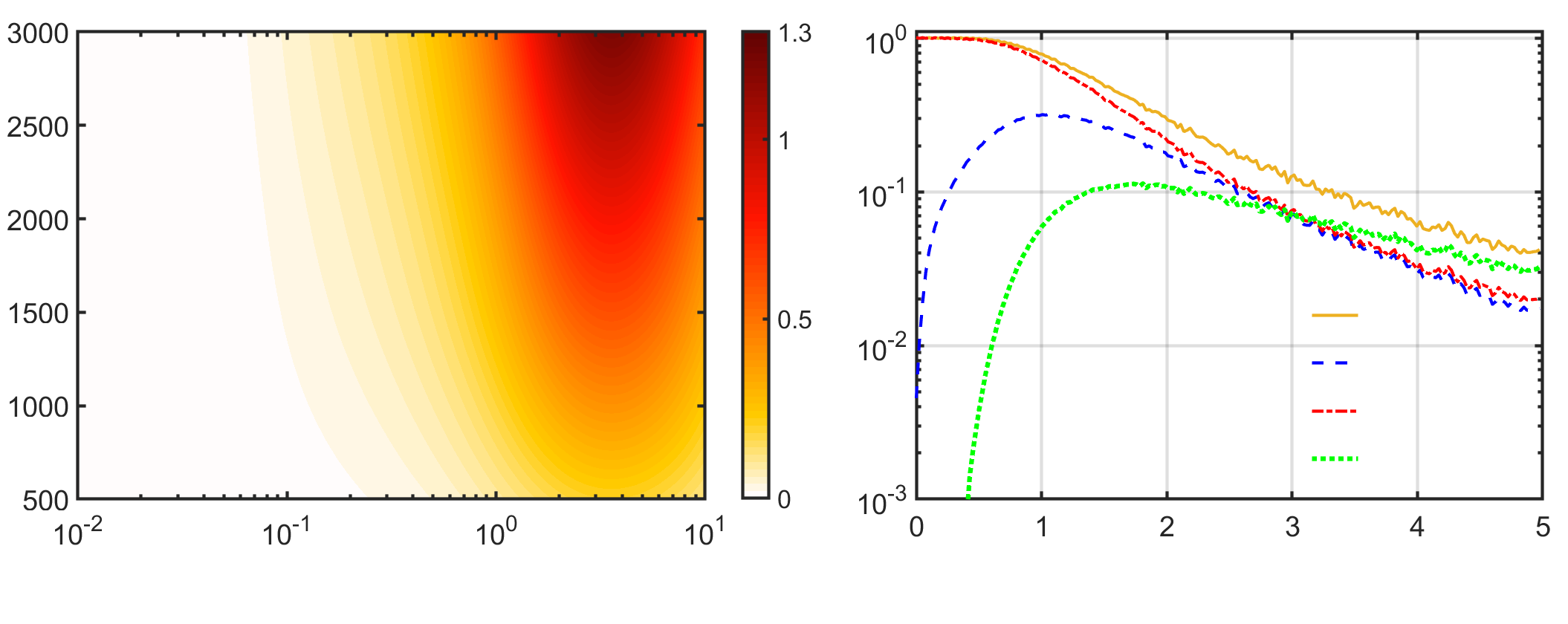}\\
	 \begin{picture}(300,20)    
    \put(17,104){(a)}
    \put(60,22){$\chi$}
    \put(-7,70){\rotatebox{90}{\small$a_0 N$}}
    \put(229,104){(b)}
    \put(190,22){$\chi$}
    \put(126,75){\rotatebox{90}{\small$S_i$}}
    \put(216,67){\fontsize{7pt}{\baselineskip}\selectfont $S_0$}
    \put(216,59){\fontsize{7pt}{\baselineskip}\selectfont $S_1$}
    \put(215,51){\fontsize{7pt}{\baselineskip}\selectfont $|S_2|$}
    \put(216,43){\fontsize{7pt}{\baselineskip}\selectfont $S_3$}
	\end{picture}\\
	 \caption{ (a) Plot of $\delta\phi$ as a function of $\chi$ and $a_0 N $ for a rectangular pulse profile. (b) Final Stokes parameters for gamma photons  propagating through an ELI-NP 10 PW laser pulse $(S^{(0)}={1,0,-1,0})$. The stokes parameters are obtained by averaging over the probe photon number. }
\end{figure}
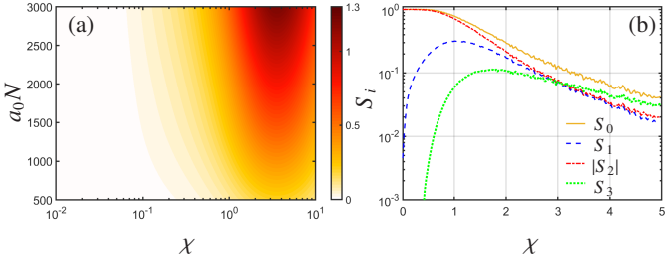

 \section{Estimate of the photon yield in the Compton process} \label{A3}

The yield of photons can be estimated using the perturbative QED theory for linear Compton scattering \cite{berestetskii1982quantum}. The total cross section for photons scatted by angles $\varphi\in\left[0,2\pi\right]$ and $\theta\in\left[0,\theta_{max}\right]$ is 
\begin{equation}
\sigma_{bs}=\frac{4\pi r_{e}^{2}}{m^{2}x^{2}}\int_{0}^{\theta_{max}}\omega_\gamma^{2}F_{0}\sin\theta,
\end{equation}
where $r_{e}=\alpha/m=2.818\times10^{-13}$cm with $m$ being the electron mass and
\begin{align*}
F_{0}	&=V+U^{2}+2U,\\
V	&=x/y+y/x,U=2/x-2/y,\\
x	&=\frac{2pk_{0}}{m^{2}}=\frac{2\varepsilon\omega_{0}}{m^{2}}\left(1+\beta\right),\\
y	&=\frac{2pk'}{m^{2}}=\frac{2\varepsilon\omega_\gamma}{m^{2}}\left(1-\beta\cos\theta\right),\numberthis
\end{align*}
$\beta=\left|\vec{p}\right|/\varepsilon_0$ with $\varepsilon_0=8.4$GeV being initial electron energy and $\vec{p}=\varepsilon_0/m$ the electron momentum, $\omega_{0}=1.55$eV the energy of laser photon for linear Compton scattering. The energy $\omega_\gamma$ of the final photon is determined via four-momentum conservation and is given by 
\begin{equation}
\omega_\gamma=\frac{\left(1+\beta\right)\varepsilon\omega_{0}}{\varepsilon+\omega_{0}-\left(\varepsilon\beta-\omega_{0}\right)\cos\theta}.
\end{equation}
For $\theta_{max}=0.05$mrad, we have $\sigma_{bs}\approx2.68r_{e}^{2}$. Employing pulse duration $\Delta t=10$ps, $I=4.3\times10^{16}\textrm{W/c\ensuremath{\textrm{m}^{2}}}$, $\omega_{0}=1.55$eV, one could estimate the yield of gamma photons via $N_{\gamma}/N_{e}=\sigma_{bs}\left(I/\omega_{0}\right)\Delta t\approx0.4$, which is roughly coincide with our simulation results.  

\section{M{\"u}ller polarimetry for detecting photon polarization}\label{A4}

When measuring vacuum birefringence via photonic signals, previous approaches employed small $\chi_\gamma$ or short interaction length to mitigate background noise stemming from real pair production. Consequently, the acquired ellipticity by probe photons was typically too small for detection. However, our method utilizes larger $\chi_\gamma$, leading to the remaining probe photons acquiring substantial circular polarization. This significant enhancement enables the measurement of vacuum polarization using photonic signals. Note however, that the pair production in this regime significantly suppresses the number of survived photons, having impact on the accuracy of the measurement.  
Even though the gamma-ray polarimetry for circular polarization poses challenges, the decrease of $\xi_1$ and increase of $\xi_3$ can be regarded as the photonic signals for detecting VB and VD, respectively. The polarization of gamma photons can also be detected by converting photons to electron and positron pairs in a high Z target. The asymmetry of the angular distribution of produced pairs can be used as photonic signals of vacuum birefringence and dichroism \cite{bragin2017high}. 

The cross section of electron-positron photoproduction by a photon with energy $\omega\ll m$ colliding with an atom (charge number $Z$) is given by \cite{bragin2017high,olsen1959photon}
\begin{align}
d\sigma_{pp} & =\frac{d\varphi}{2\pi}\left\{ \sigma_{0}S_{0}+\sigma_{1}\left[S_{1}\sin2\varphi+S_{3}\cos2\varphi\right]\right\},
\end{align}
where
\begin{align*}
\sigma_{0}&=2\frac{Z^{2}\alpha r_{e}^{2}}{\omega^{3}}\int_{m}^{\omega-m}d\varepsilon\int_{m^{2}/\varepsilon^{2}}^{1}d\zeta\left\{ \left(\varepsilon^{2}+\varepsilon'^{2}\right)\left(3+2\varGamma\right)\right.\\&+\left.2\varepsilon\varepsilon'\left[1+4u^{2}\zeta^{2}\varGamma\right]\right\} ,\\\sigma_{1}&=2\frac{Z^{2}\alpha r_{e}^{2}}{\omega^{3}}\int_{m}^{\omega-m}d\varepsilon\int_{m^{2}/\varepsilon^{2}}^{1}d\zeta\left\{ 8\varepsilon\varepsilon'u^{2}\zeta^{2}\varGamma\right\} ,\\\varGamma&=\ln\left(1/\delta_{i}\right)-2-f\left(Z\right)+F\left(\delta/\zeta\right),\\\zeta&=1/\left(1+\vec{u}^{2}\right),\\\vec{u}&=\vec{p}_{\perp}/m=\left|p_{\perp}\right|\left(\cos\varphi,\sin\varphi\right),\\\delta&=m\omega/\left(2\varepsilon\varepsilon'\right),\\f\left(Z\right)&=\left(Z\alpha\right)^{2}\sum_{n=1}^{\infty}\frac{1}{n\left(n^{2}+\left(Z\alpha\right)^{2}\right)},\\F\left(\delta/\zeta\right)&=-\frac{1}{2}\sum_{i=1}^{3}\alpha_{i}^{2}\ln\left(1+B_{i}\right)\\&+\sum_{i,j=1,i\neq j}^{3}\alpha_{i}\alpha_{j}\left[\frac{1+B_{j}}{B_{i}-B_{j}}\ln\left(1+B_{j}\right)+\frac{1}{2}\right],\\B_{i}&=\left(\beta_{i}\zeta/\delta\right)^{2},\beta_{i}=\left(Z^{1/3}/121\right)b_{i},\\\alpha_{1}&=0.1,\alpha_{2}=0.55,\alpha_{3}=0.35;\\b_{1}&=6,b_{2}=1.2,b_{3}=0.3.
\end{align*}
For photons with a wide spectrum [see Fig. \ref{Fig.pho1d} (a)], the asymmetries for detecting vacuum birefringence ($R_{B}$) and vacuum dichroism ($R_{D}$) are
\begin{align*}\label{RBD}
R_{B}&=\frac{\left(N_{\pi/4}+N_{5\pi/4}\right)-\left(N_{3\pi/4}+N_{7\pi/4}\right)}{\left(N_{\pi/4}+N_{5\pi/4}\right)+\left(N_{3\pi/4}+N_{7\pi/4}\right)},\\
R_{D}&=\frac{\left(N_{0}+N_{\pi}\right)-\left(N_{\pi/2}+N_{3\pi/2}\right)}{\left(N_{0}+N_{\pi}\right)+\left(N_{\pi/2}+N_{3\pi/2}\right)},\numberthis
\end{align*}
where $N_{\beta_0}$ denotes the number of pairs detected in the azimuthal angle range $\varphi\in(\beta_0-\beta,\beta_0+\beta)$ of the transverse plane, with $\beta=33^\circ$ as optimal angle for both observables, and
\begin{align*}
N_{\pi/4}&=N_{5\pi/4}=\sum_{i}N_{\gamma i}n_{z}l\int_{\pi/4-\beta}^{\pi/4+\beta}\frac{d\varphi}{2\pi}\left\{ \sigma_{0i}S_{0i}\right.\\&+\left.\sigma_{1i}\left[S_{1i}\sin2\varphi+S_{3i}\cos2\varphi\right]\right\} \\&=\sum_{i}N_{\gamma i}n_{z}l\left[\sigma_{0i}S_{0i}\frac{\beta}{\pi}+\sigma_{1i}S_{1i}\frac{\sin2\beta}{2\pi}\right],\\N_{3\pi/4}&=N_{7\pi/4}=\sum_{i}N_{\gamma i}n_{z}l\left[\sigma_{0i}S_{0i}\frac{\beta}{\pi}-\sigma_{1i}S_{1i}\frac{\sin2\beta}{2\pi}\right],\\N_{0}&=N_{\pi}=\sum_{i}N_{\gamma i}n_{z}l\left[\sigma_{0i}S_{0i}\frac{\beta}{\pi}+\sigma_{1i}S_{3i}\frac{\sin2\beta}{2\pi}\right],\\N_{\pi/2}&=N_{3\pi/2}=\sum_{i}N_{\gamma i}n_{z}l\left[\sigma_{0i}S_{0i}\frac{\beta}{\pi}-\sigma_{1i}S_{3i}\frac{\sin2\beta}{2\pi}\right].\numberthis
\end{align*}
Here, the subscript $i$ denotes the variables for photons with energy of $\omega_i$. Substituting the above expressions of $N_{\beta_0}$ into Eq. (\ref{RBD}), we have
\begin{align*}
R_{B}&=\frac{\sum_{i}N_{\gamma i}\sigma_{1i}S_{1i}\sin\left(2\beta\right)}{\sum_{i}N_{\gamma i}\sigma_{0i}S_{0i}2\beta},\\
R_{D}&=\frac{\sum_{i}N_{\gamma i}\sigma_{1i}S_{3i}\sin\left(2\beta\right)}{\sum_{i}N_{\gamma i}\sigma_{0i}S_{0i}2\beta}.\numberthis
\end{align*}
Using a effective thickness of $3.66\times10^{20}$ corresponding to a conversion efficiency of $\eta=0.01$  with $\sigma_0=344r_e^2$ in \cite{bragin2017high}, and the polarization distribution of photon after interaction with the laser [see Fig.6 (a)],  we obtained the asymmetry $R_B=-0.0369$ and $R_D=0.0246$. In the case the photons do not interact with the laser, we obtain $R_B^0=-0.0617$ and $R_D^0=0$. The observables of vacuum polarization are ${\cal A}_{B}=R_{B}-R^0_{B}=0.0247$ and ${\cal A}_{D}=R_{D}-R^0_{D}=0.0246$.

The produced pair number is
\begin{align}
N_{e^{+}e^{-}} & =\sum_{i}N_{\gamma i}n_{z}l\sigma_{0i}S_{0i}\frac{4\beta}{\pi},
\end{align}
which gives the standard deviation $\Delta R_{B,D}=1/\sqrt{N_{e^{+}e^{-}}}$.
Assuming an electron bunch with $N_{e^-}^0=1\times 10^8$ is used for linear Compton scattering, the photon yield within 0.05~mrad is around $N_{\gamma}^0=0.5\times 10^7$. 
Then the confidence level for a single short could reach  
\begin{align*}
n_{B} & =\frac{{\cal A}_{B}}{\Delta R_B}={\cal A}_{B}\sqrt{\frac{4\beta n_{z}l}{\pi}\sum_{i}N_{\gamma i}\sigma_{0i}S_{0i}}=8,\\
n_{D} & =\frac{{\cal A}_{D}}{\Delta R_D}={\cal A}_{D}\sqrt{\frac{4\beta n_{z}l}{\pi}\sum_{i}N_{\gamma i}\sigma_{0i}S_{0i}}=8,\numberthis
\end{align*}
indicating a single short measurement of vacuum polarization could reach a confidence level of $8\sigma$.

Note that, one should make sure that the observables ${\cal A}_{B}=0.0247$ are much larger than the error of the initial photon polarization measurement $\sim \Delta R_B^0=1/\sqrt{N^0_{e^+e^-}}=0.0035$ as $\xi_{10}\neq 0$ for initial gamma-rays. Apparently, the condition ${\cal A}_{B}-\Delta R_B^0>\Delta R_B^0$ is fulfilled for a single shot.
However, as for previous schemes \cite{bragin2017high}, the feasibility relies on the capacity of post-selection techniques to reduce the substantial background noise from radiation and cascaded detectors to enhance the conversion efficiency \cite{tavani2003agile,atwood2009large,peitzmann2013prototype}  and suppress multiple Coulomb scattering \cite{gros2017gamma}. We emphasize that the experimental detection capacity for gamma polarization (typically $\gtrsim$10\% \cite{ozaki2016demonstration})  is currently significantly lower than that for positrons  (typically $\sim$ 0.5\% \cite{narayan2016precision}).

}

\bibliography{reference}

\end{document}